\documentclass[letterpaper,twocolumn,10pt]{article}
\usepackage{usenix,epsfig,endnotes}
\usepackage{cite}
\usepackage{graphicx}
\usepackage{array}
\usepackage{mdwmath}
\usepackage{mdwtab}
\usepackage{eqparbox}
\usepackage[font=footnotesize,caption=true]{subfig}
\usepackage{url}
\usepackage[normalem]{ulem}
\usepackage{comment}
\usepackage[pdftex,colorlinks=true,citecolor=black,filecolor=black,
linkcolor=black,urlcolor=black]{hyperref}
\usepackage{times}
\usepackage{listings}

\usepackage{enumitem}
\usepackage{tabularx}
\usepackage{amssymb}
\usepackage{xspace}
\usepackage{framed}
\usepackage[usenames,dvipsnames]{color}
\usepackage{makecell}
\usepackage{ulem}
\usepackage{flushend}
\newcommand{\myparagraph}[1]{\vspace{0.25em}\noindent\textbf{#1:}}
\newcommand{\myparagraphs}[1]{\noindent\textbf{#1:}}
\usepackage{dsfont}
\newcommand{\user}[1]{{$\mathds{P}#1$}}

\newcommand{\pbe}{PyBE\xspace}
\newcommand{\pbes}{PyBE\xspace}

\newcommand{\emparagraph}[1]{\vspace{0.25em}\noindent\textit{#1}}

\definecolor{dkgreen}{rgb}{0,0.6,0}
\definecolor{gray}{rgb}{0.5,0.5,0.5}
\definecolor{mauve}{rgb}{0.58,0,0.82}
\definecolor{shadecolor}{rgb}{0.95,0.95,0.95}
\usepackage{array}
\newcolumntype{L}{>{\centering\arraybackslash}m{5cm}}
\excludecomment{techreport}
\includecomment{conference}

\begin{document}
\title{Policy by Example: An Approach for Security Policy Specification}
\author{
{\rm Adwait Nadkarni}\thanks{This author was a student at the North Carolina State University at the time this work was completed.}\\
apnadkarni@wm.edu\\
The College of William and Mary
\and
{\rm William Enck}\\
whenck@ncsu.edu\\
North Carolina State University
\and
{\rm Somesh Jha}\\
jha@cs.wisc.edu\\
University of Wisconsin-Madison
\and
{\rm Jessica Staddon}\\
jessica.staddon@ncsu.edu\\
North Carolina State University
}
\date{}
\maketitle
\thispagestyle{plain}
\pagestyle{plain}

\begin{abstract}
Policy specification for personal user data is a hard problem, as it
depends on many factors that cannot be predetermined by system
developers. Simultaneously, systems are increasingly relying on users to
make security decisions.
In this paper, we propose the approach of Policy by Example (\pbe) for
specifying user-specific security policies. \pbes brings the benefits of
the successful approach of programming by example (PBE) for program
synthesis to the policy specification domain. In \pbe, users provide
policy examples that specify if actions should be allowed or denied in
certain scenarios.  \pbe then predicts policy decisions for new
scenarios. 
%
%
A key aspect of \pbe is its use of active learning to enable users to
correct potential errors in their policy specification.
To evaluate \pbe's effectiveness, we perform a feasibility study with
expert users.
Our study demonstrates that \pbe correctly predicts policies with 76\%
accuracy across all users, a significant improvement over naive
approaches. 
Finally, we investigate the causes of inaccurate predictions to motivate
directions for future research in this promising new domain.
%
%
\end{abstract}

\vspace{-0.5em}
\section{Introduction}
\label{sec:intro}
\vspace{-0.5em}

In the era of pervasive computing, the security of user data and
resources is of paramount importance.  Complex systems such as IoT
platforms (e.g., IFTTT~\cite{tt10} and SmartThings~\cite{smartthings}),
smartphone platforms (e.g., Android and iOS) and even traditional
commodity platforms are being leveraged for processing user data.
However, our knowledge of policy specification has not kept pace with
the rise of complex systems that are increasingly relying on the user to
specify the security policy.

Further, user data has become increasingly user-specific.  Users no
longer directly deal with generic files, but create specific data
objects such as notes, whiteboard snapshots, and selfies. This data is
abstract, i.e., its importance and properties are subjective.  System
designers or application developers cannot specify a security policy
for abstract user data. The situation is even critical for novel
security systems that provide strong data security guarantees for
user data (e.g., decentralized information flow control (DIFC)
systems for Android~\cite{ne13,jaf+13,xw15,naej16},
Chromium~\cite{bcj+15}). Such systems are impractical to deploy unless
users specify security policies; and users are bad at specifying
security policies~\cite{shc+09,mr05} without assistance.

This paper raises the simple but important question of policy
specification: how to teach the system {\em what} the user wants to
protect, and {\em how} the user wants to protect it?  Consider the
following example: a smartphone user wants to synchronize all personal
notes with her cloud account, except notes labeled as medical data.
Since we are dealing with user-specific data-use scenarios, we can
justifiably expect the user to provide some input to the system.
However, expecting the user to enumerate every possible scenario
involving medical data is impractical. The policy must be predicted.

We propose the approach of specifying Policy by Example (\pbe) for
user-specific data. \pbe is inspired by the successful use of
programming by example (PBE) for program synthesis. Specifically, we
emulate the approach of Gulwani~\cite{gul11}, where the user specifies
examples consisting of the input and output, and the system learns a
program that can predict the output for unknown (but similar) inputs.
Similarly, in \pbe, the user specifies policy examples, in terms of the
data-use scenario (i.e., the input) and the policy decision (i.e., the
output).  The system uses the policy examples to predict policy
decisions for new scenarios. By requiring only relevant examples, and
not complete policy specification, \pbe makes policy specification
tractable.

Predicting security policies for abstract, user-specific data with
unknown properties is hard, as the learner cannot make any assumptions
about the input data points. In contrast, prior work on predicting
privacy policies for well-known private data~\cite{khsc08,cms11} can
make assumptions that aid prediction; e.g., Cranshaw et al.~\cite{cms11}
take advantage of probabilistic models to learn location privacy
policies knowing that location and time are continuous variables. \pbe
cannot make any such assumptions, which puts us at a significant
disadvantage.  However, this disadvantage drove us to embrace a simpler
approach that does not demand specific properties from data.

We chose a variant of the {\it $k$ nearest neighbor (kNN)}
classifier~\cite{mur12} for predicting policies. Our key requirements
were that the algorithm be {\sf (1)} non-parametric, i.e.,
independent of models that rely on fixed set of parameters, and {\sf
(2)} easy to explain, i.e., for the user to understand how the policy
was inferred. Recall that a policy example is composed of a scenario and
the corresponding policy decision. To predict the policy decision for a
new scenario, our algorithm performs a nearest neighbor search for
finding similar scenarios from the user's examples, and predicts the
majority policy decision. 

An important challenge in applying kNN is calculating the distance
between data points.  To calculate distance between scenarios, we treat
scenarios as Boolean functions, and propose a novel distance metric for
the same. As some policies may be relatively more important to
the user, we extend our metric to support weights. Note that existing
distance metrics (e.g., {\em jaccard} distance) may require significant
re-engineering to incorporate weights, which motivates our development
of a new metric.

\pbe recognizes that policy specification by users in any form is error
prone.  A key contribution is our use of active learning for enabling
the user to correct policy decisions. We draw inspiration from the work
of Gulwani~\cite{gul11}, which detects noise in the user's examples, and
prompts the user for new outputs for problematic examples. Similarly,
\pbe uses noise in the user's policy examples as an indication of error
in policy decisions, and engage the user in correcting errors. 

We evaluate the feasibility of \pbe with a study of expert users. Our
study involves 8 participants, and 5 target security policies (e.g.,
exporting to the enterprise cloud), i.e., we solve 40 independent policy
specification problems. Our participants generate 246 policy scenarios
in total, and assign decisions for the 5 policies,
resulting in a total of 1,230 policy examples across participants. 

We perform two experiments with this data.  First, we find errors in
policy decisions using a manual review and a \pbe-assisted interactive
review of policy examples. Then, we test \pbe's prediction for randomly
generated scenarios with unknown policy decisions.  \pbe demonstrates a
prediction accuracy of over 76\% across all participants, and fares
better than our assumed baseline of a random coin flip, and a naive
approach.  A significant finding is that the \pbe-assisted interactive
review approach helped participants find {\em five times} as many errors
as their manual reviews.

Our evaluation is evidence of the feasibility, i.e., the effectiveness
of PyBE in terms of both prediction accuracy and error identification,
but does not speak to the general usability of PyBE.  Although 8
participants is small for a human study, the evaluation is able to
answer important questions through the analysis of user-generated policy
examples (i.e., 1,230 user-generated examples). The research questions
answered in the evaluation operate at the level of policy examples,
making the dataset sufficiently large for evaluating feasibility.

The contributions of this paper are as follows:
\begin{itemize}\renewcommand{\itemsep}{-0.3em}
  \item We introduce the Policy by Example (\pbe) paradigm for
    predicting user-specific security policies. Our approach takes
    labeled policy scenarios from the user, and predicts policy
    decisions for new policy scenarios.
  \item We use an interactive approach to assist users in finding
    incorrect policy decisions in their examples. We empirically
    demonstrate its effectiveness over manual policy
    reviews.  
  \item We perform a feasibility study with expert users, and
    demonstrate better prediction accuracy than both a baseline as well
    as a naive approach.  
\end{itemize}

This paper is the first step in our vision of a policy assistant for
user data. With \pbe, we provide an approach for predicting security
policies for user-specific data, and demonstrate its technical
feasibility. Further, we analyze our incorrect predictions, and describe
the lessons we learned in the process.  Finally, we describe challenges
(e.g., usability for non-experts, modeling policy change) and future
research directions in this promising new area.

\vspace{-0.5em}
\section{Related Work}
\label{sec:relwork}
\vspace{-0.5em}

The notion of Policy by Example (\pbe) is inspired by recent work in the
domain of Programming by Example (PBE). The objective of PBE is simple:
if the user knows the steps for performing a task, the user should not
have to write a program; instead, the computer should learn from the
user's actions on an example, and generalize the
program~\cite{ch93,lie01}. However, the user may not always be able to
express the reasoning, or the intermediate steps, involved in creating a
program.  Recent work by Gulwani~\cite{gul11} makes PBE feasible
for such programming tasks, by using only input-output examples to
synthesize a program that predicts outputs for unseen inputs. \pbe
follows a similar intuition, and predicts policy decisions for new
scenarios using only input-output examples (i.e., policy scenarios and
corresponding decisions).

However, \pbe does not generalize the program before testing, as is
often done in PBE.  That is, while the proposed paradigm is conceptually
similar to PBE, the process used to predict policies borrows from
another well-established domain: case-based reasoning
(CBR)~\cite{kol93}. In CBR, the outcome of a test case is determined by
looking at the outcomes of previously observed cases (e.g.,
legal reasoning using precedents). In a way, CBR mimics a human expert's
reasoning, and performs lazy generalization of domain knowledge at
testing time. CBR has been successfully used in many domains, e.g.,
synthesizing music~\cite{ads98,ad01}, providing decision support in
molecular biology~\cite{jrg+01},  and for solving spatial reasoning
problems~\cite{hol99}. However, to our knowledge, CBR has never been
used for predicting user security policies, and \pbe is novel in its use
of a similarity heuristic (i.e., a form of CBR) for predicting security
policies.

A critical advantage of CBR is that it provides a way to deal with
uncertainty, in contrast with the process of eager learning (e.g., rule
induction). Prior user-controllable methods for predicting privacy
policies for well-known private data (e.g., Location) use eager
learning, which requires making strategic parameter choices for
generalization, often based on some known properties of the training
data~\cite{khsc08,fl10,cms11}.  For example, Cranshaw et
al.~\cite{cms11} use a probabilistic model to learn location privacy
policies, assuming the availability of a large number of data points
since location is a continuous variable.  However, \pbe cannot make such
assumptions for user data with uncertain properties (e.g., Bob's scanned
documents, Alice's notes), and uses a form of CBR, which does not
require a priori generalization.

Prior work has proposed usable interfaces for eliciting security
responses, which are relevant for our long-term vision of creating a
policy assistant for user data.  For instance, a prototype of \pbe for a
computing device may adopt the ``interactive dropdowns'' in Johnson et
al.'s interactive policy authoring template for specifying initial
examples~\cite{jkkg10a,jkkg10b}. Similarly, Reeder et al's ``expandable
grids'' may be adapted for visualizing policy examples for the
user~\cite{rbc+08}. Such work only provides interfaces, and does not
fulfill \pbe's objective of making policy specification feasible through
prediction.  Further, recent work on user-driven access control  (e.g.,
Roesner et al.~\cite{rkm+12}, Ringer et al.~\cite{rgr16}) provides a
usable way of acquiring the user's policy decision, by embedding access
permissions into the user's natural UI flow of accessing resources.
However, defining specific permissions (i.e., gadgets) for an
exponential space of subjective and user-specific data-use scenarios may
be infeasible.

Prior work also complements the specification of user-specific
policies, by providing content recognition for automatically tagging
data for \pbe~\cite{vscy09,sslw11,slb13,bbzl15,slsw15}, or by providing
security profiles for standard, well-known, security settings (e.g.,
Android permissions)~\cite{lls14,llsh14}, allowing \pbe to focus on
predicting policies for abstract, user-specific data. 

Finally, while \pbe assists the user in specifying policies for
user-specific data, there has been prior research in the domain of
policy specification to help application or system developers.  Prior
work provides application developers with tools for expressing their
security policies~\cite{ek08,he08,hjr10,sxwx14}.  Further, in contrast
with prior work that assists developers in expressing known policies,
Slankas et al. aid the developer by extracting access control rules from
application-specific text artifacts using natural language processing
(NLP)~\cite{sxwx14}.  Similarly, access control logs and system call
traces have previously been used to refine the system's security
policies (e.g., EASEAndroid~\cite{wer+15} and Polgen~\cite{shr06}).

\vspace{-1em}
\section{Motivation and Problem}
\label{sec:motivation}
\vspace{-1em}

User data and data-use scenarios are user-specific. External observers
such as system designers or application developers cannot specify the
user's security policy without knowing the user's context of data
use~\cite{nis04,bdmn06}. Moreover, this constraint is not limited to
user-owned data; prior work demonstrates that even the security
preferences for enterprise data vary with users and personal data-use
contexts~\cite{gs10}. Consider the following example, which describes
how two users may differ in terms of the relevance of data-use scenarios
as well as security preferences for the same scenarios.

\myparagraph{Example} Alice and Bob are two smartphone users, who use a
fictional note-taking application {\em Notes} (similar to Google Keep)
on their smartphones to collect and organize information. {\em Notes}
backs up data to a designated cloud provider (e.g., Google Drive).
Alice consolidates expenses by scanning paper receipts into the {\em
Notes} application. However, Alice does not trust the cloud with medical
data, and wants medical receipts (i.e., receipts scanned at the
hospital) to only be stored locally, and not synced. Similarly, Bob uses
{\em Notes} to aggregate his documents. As {\em Notes} is set up with
Bob's enterprise cloud, he does not wish to sync personal documents
(e.g., documents created after work hours). That is, the requirements
for {\em what} users want to protect (i.e., relevant data-use scenarios)
are user-specific.

Further, even when two users may agree on {\em what} they want to
protect, they may not agree on {\em how} they want to protect it.
Suppose Alice and Bob meet at a conference and exchange business
cards. Alice is self-employed, and feels confident in backing up
business cards acquired after work hours to her enterprise cloud.
However, Bob does not want to disclose networking opportunities to
his company by syncing cards collected after work hours
to his enterprise cloud. Security preferences for user data stem from
the user's personal circumstances.

\myparagraph{Problem} 
In this paper, we focus on the problem of specifying user-specific
security policies.  The nature of the problem dictates that the policy
specification must receive input from the user. However, it is
impractical to expect the user to specify the policy for every scenario
in an exponential space.  Hence, this paper addresses the problem of
predicting the security policy for new data-use scenarios, based on the
scenarios previously described by the user. 

\vspace{-1em}
\section{Policy by Example (\pbe)} 
\label{sec:pbe}
\vspace{-0.5em}

\begin{figure}[t]
  \centering
  \subfloat[][The user specifies examples. \pbe suggests
  error-correction.]{\includegraphics[width=1.2in]{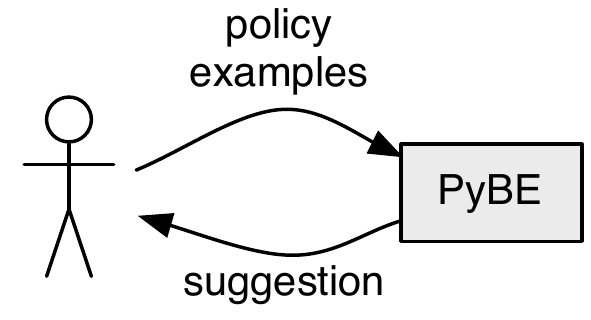}\label{fig:overview_a}}\qquad
  \subfloat[][\pbe predicts policy decisions for new
  scenarios.]{\includegraphics[width=1.4in]{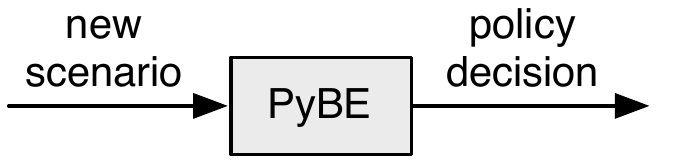}\label{fig:overview_b}}
  \vspace{-1em}
  \caption{An overview of the \pbe approach.}
  \label{fig:overview}
  \vspace{-0.5em}
\end{figure}

\pbe is inspired by recent work on Programming by Examples~\cite{gul11},
which learns a program from input-output examples. As shown in
Figure~\ref{fig:overview_a}, the user provides policy examples (i.e.,
data-use scenarios and policy decisions), and \pbe interactively
suggests corrections to the user's policy decisions.  \pbe then predicts
policy decisions for new scenarios as shown in
Figure~\ref{fig:overview_b}.

This section provides the intuition behind our approach. We describe
\pbe formally in Section~\ref{sec:algorithm}.  We start by describing
the structure of a policy example.

\vspace{-1em}
\subsection{The Policy Example} 
\label{sec:policy-example}
\vspace{-0.5em}
A policy example is composed of a {\em scenario}, and a {\em policy
decision} (i.e., {\sf allow}/1 or {\sf deny}/0) for that scenario. A
scenario is as a set of {\em tags}, where each tag denotes the resource
to be protected (e.g., business card) or a condition that influences the
policy (e.g., created after work hours). Using a set of tags enables
users to describe complex scenarios composed multiple conditions or data
objects. Our use of tags is motivated by prior work that
demonstrates that users can effectively re-purpose organizational tags
to express access control policies~\cite{klms+12}. 

In addition to the user-customizable policy example, we also define a
fixed policy {\em target} which represents the action controlled by the
policy; e.g., exporting data to the enterprise cloud, i.e., the {\em
WorkCloud} policy target. Policy specification is performed separately
for each policy target, i.e., {\em independent} of other targets. Thus, each
target represents a separate high-level policy that must be specified
(e.g., the user's {\em WorkCloud} policy).  The policy targets used in
this paper are motivated by prior work on restricting the network export
of secret data~\cite{sym+14,bcj+15,naej16}.

\begin{table}[t]
\footnotesize
\centering
\caption{Bob's examples for the {\em WorkCloud} policy.}
\vspace{-1em}
\label{tbl:bob-policy}
\begin{tabular}{l|c|c}
  \Xhline{2\arrayrulewidth}
  {\bf No.} & {\bf Scenario}             & {\bf Policy Decision}\\
  \Xhline{2\arrayrulewidth}
  1 & \{Home, Photo\}   & deny  \\  
  2 & \{Work, Photo\}   & allow  \\  
  3 & \{Document\} & allow \\
  \Xhline{2\arrayrulewidth}
    \end{tabular} 
    \vspace{-1.5em}
\end{table} 

Table~\ref{tbl:bob-policy} shows Bob's policy examples for the {\em
WorkCloud} policy target. We describe each example, along with Bob's
security requirement behind it.
First, Bob considers data created at home to be personal, so Bob's
photos created at home must never be exported to the enterprise cloud.
Thus, Bob denies export for example~1, i.e, \{{\tt Home,Photos}\}.
Second, photos taken at work may be exported to the enterprise cloud.
Hence, Bob allows export for example~2, i.e., \{{\tt Work,Photos}\}.
Third, Bob does not (currently) imagine a situation where he would deny
export for documents.  Hence, Bob allows export for example~3, i.e.,
\{{\tt Document}\}.  We use Bob's examples to describe
\pbe. 

\vspace{-1em}
\subsection{Our Approach}
\label{sec:overview}
\vspace{-0.5em}

As described previously, \pbe uses a variation of the kNN algorithm for
predicting policies. That is, Bob provides \pbe with a set of policy
examples (i.e., scenarios labeled with policy decisions). When faced
with a new scenario with an unknown policy decision, we perform a
nearest neighbor search of Bob's examples. That is, we search Bob's
examples for the closest examples, i.e., examples with scenarios closest
to the new scenario, and predict the policy decision of the majority of
the closest examples.  Note that distance between examples is described
in terms of their scenarios (i.e., when we say ``examples are close'',
it means their scenarios are close). 

An approach for predicting security policies should be deterministic if
we want users to understand its outcome (i.e., independent of arbitrary
parameters).  Based on this rationale, we eliminate the need to specify
the parameter $k$. Our variation of kNN considers the {\em closest}
neighbors as all neighbors at the closest distance, instead of k
neighbors at varying distances.

We now demonstrate our approach with a manual walk-through. A manual
walk-through is feasible because the basic process of kNN is intuitive
and its outcome is easy to explain. 
To demonstrate our approach, we predict
policy decisions for the following new scenarios for Bob: \{{\tt Home}\}
and \{{\tt Home,Document}\}, using Bob's initial policy specification
shown in Table~\ref{tbl:bob-policy}.

Consider the first new scenario, \{{\tt Home}\}.  Just by looking at
Bob's specification in Table~\ref{tbl:bob-policy}, the reader may
identify example~1 (i.e., \{{\tt Home,Photo}\}) as closest to the new
scenario, since it is the only example that includes the tag {\tt Home}.
As a result, we predict the policy decision for the new scenario
{\tt \{Home\}} as {\sf deny}, i.e., as the decision of its nearest
neighbor \{{\tt Home,Photo}\}. This decision mirrors Bob's assumption of
data created at home being personal, and not exportable to the
enterprise cloud. \pbe's distance metric described in
Section~\ref{sec:algorithm} uses a similar property for computing
distance between two examples, and comes to the same conclusion. 

Now consider the second new scenario, \{{\tt Home,Document}\}. This
time, there are two examples that seem to be equally close to the new
scenario, i.e., \{{\tt Home,Photo}\} and \{{\tt Document}\}, since they
each
have one tag in common with \{{\tt Home,Document}\}. Since both the
nearest examples have different policy decisions, our simple metric is
insufficient. This is one of the motivations for introducing
weights.  Suppose Bob considers personal data created at home (i.e., the
tag {\tt Home}) to be most confidential. Therefore, Bob assigns {\tt
Home} more ``importance'' (i.e., a higher weight) than any other tag in
terms of its influence on the policy decision.  As a result, the new
scenario \{{\tt Home,Document}\} can be deemed closer to \{{\tt
Home,Photo}\} than \{{\tt Document}\}, as {\tt Home} has a higher weight
and more say in the decision than the other tags, e.g.,
{\tt Document}. Thus, export is denied for \{{\tt Home,Document}\},
which aligns with Bob's preference of data created at home being
personal, and not exportable to the enterprise cloud. 

The purpose of weights is not limited to breaking ties. Suppose Bob
specifies another example, i.e., \{{\tt Document,Receipt}\}, with
decision {\sf allow}. Now consider another new scenario \{{\tt
Document,Receipt,Home}\}. Without any knowledge of weights, it is easy
to see that \{{\tt Document,Receipt}\} would be the example closest to
the new scenario \{{\tt Document,Receipt,Home}\}, resulting in {\sf
allow} being predicted (i.e., there is no tie).  At the same time, we
know that Bob has allocated a higher weight to {\tt Home}, since Bob
considers home data to be confidential and important with respect to the
{\em WorkCloud} target. The weights ensure that \{{\tt
Document,Receipt,Home}\} is closer to \{{\tt Home,Photo}\} instead of
\{{\tt Document,Receipt}\}, and export is denied as per Bob's actual
security preference. Simply stated, weights enable the user to make some
information tags beat others in the distance calculation.  Our weighted
metric described in Section~\ref{sec:weighted-metric} follows a similar
rationale.

\begin{table}[t]
\footnotesize
\centering
\caption{Bob's extended set of examples for the {\em WorkCloud} policy target, with newly added examples in {\bf bold}.}
\vspace{-1em}
\label{tbl:bob-policy-revised}
\begin{tabular}{l|c|c}
  \Xhline{2\arrayrulewidth}
  {\bf No.} & {\bf Scenario}             & {\bf Policy Decision}\\
  \Xhline{2\arrayrulewidth}
  1 & \{Home, Photo\}   & deny  \\  
  2 & \{Work, Photo\}   & allow  \\  
  3 & \{Document\} & allow \\
  {\bf 4} & \{{\bf Home, Document}\} & {\bf deny} \\
  {\bf 5} & \{{\bf Home, Memo}\} & {\bf allow} \\
  \Xhline{2\arrayrulewidth}
    \end{tabular} 
\vspace{-1em}
\end{table} 

An important contribution of \pbe is that it recognizes that policy
specification by users can be error-prone. \pbe uses active learning to
engage the user in finding and correcting potential errors in their
policy decisions.  Our approach is inspired by the work of
Gulwani~\cite{gul11}, which detects noise in the user's input-output
examples, and recommends changes to incorrect outputs.
Similarly, \pbe looks for noise in the user's examples, which may
indicate one or more incorrect policy decisions.  We use our
variant of kNN for this purpose.  Note that the objective of this task
is to engage the user in finding errors in existing examples, and not to
predict policy decisions for new examples.  We explain our approach with
the following extension to Bob's policy:

Suppose Bob adds two additional examples, i.e., \{{\tt Home,Document}\}
with decision {\sf deny}, and \{{\tt Home,Memo}\} with decision {\sf
allow}. We borrow the first example (\{{\tt Home,Document}\}) from the
previous discussion on weights. The second example shows Bob's policy
for a memo created at home.  Further, recall that {\tt Home} has a
higher weight, hence examples containing {\tt Home} will be closer to
each other than other examples not containing {\tt Home}.  Bob's
complete set of examples is shown in Table~\ref{tbl:bob-policy-revised}.

\begin{figure}[t]
    \centering
    \includegraphics[width=2in]{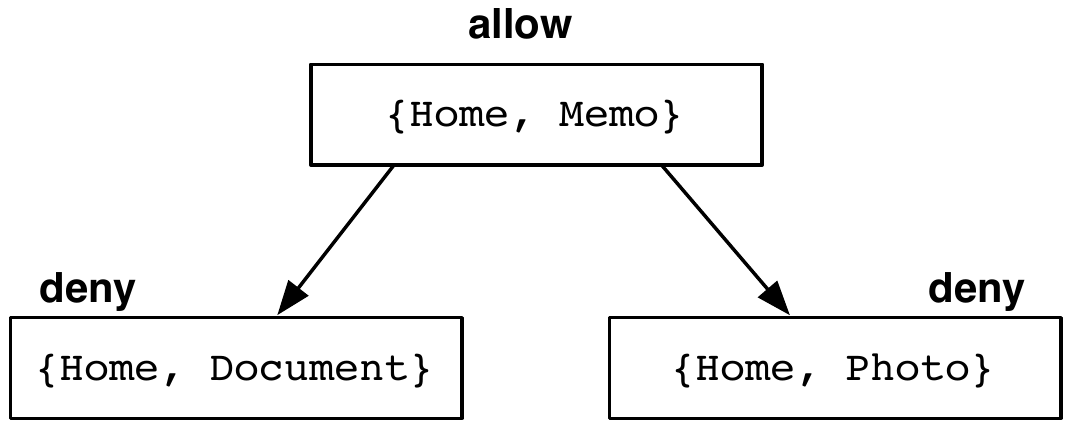}
    \vspace{-0.5em}
    \caption{\{{\tt Home,Memo}\} disagrees with the majority policy decision of its
    nearest neighbors.} 
    \label{fig:nn-graph-a}
    \vspace{-1.5em}
\end{figure}

We perform a nearest neighbor search for the example \{{\tt
Home,Memo}\}, and identify \{{\tt Home,Photo}\} and \{{\tt
Home,Document}\} as its nearest neighbors. An intuitive way of
visualizing this group of examples is in the form of a graph, such that
{\sf (1)} the examples are vertices, and {\sf (2)} directed edges are
drawn from the example for whom the search was performed to its nearest
neighbors.
Figure~\ref{fig:nn-graph-a} shows the graph for \{{\tt Home, Memo}\}.

\begin{figure}[t]
    \centering
    \includegraphics[width=2in]{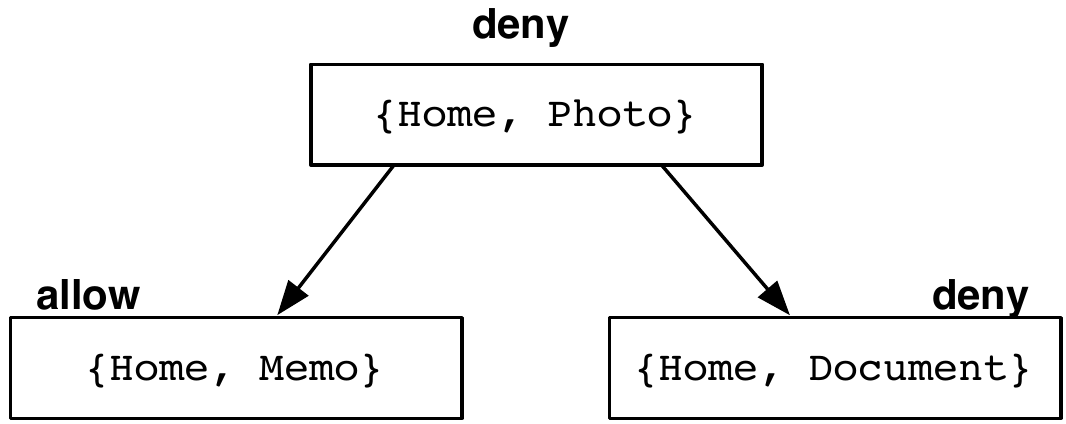}
    \vspace{-0.5em}
    \caption{
    There is no majority policy decision among the nearest neighbors of \{{\tt
    Home,Photo}\}.
    } 
    \label{fig:nn-graph-c}
    \vspace{-1.5em}
\end{figure}

If we focus on the policy decisions of the vertices in
Figure~\ref{fig:nn-graph-a}, we see that Bob's decision for \{{\tt Home,
Memo}\} (i.e., {\sf allow}) disagrees with the decision for both its
nearest neighbors. This inconsistency or noise indicates one of two
possibilities: {\sf (a)} Bob made a mistake in labeling \{{\tt Home,
Memo}\} with the decision {\sf allow}, or {\sf (b)} Bob wanted to make a
genuine exception for memos. Instead of making a guess, \pbe asks Bob.
That is, we recommend Bob to label \{{\tt Home,Memo}\} as {\sf deny} for
resolving this inconsistency, Bob may accept our recommendation, or
reject it and make an exception. Using such interactive recommendations,
\pbe engages Bob in correcting potential errors.

Figure~\ref{fig:nn-graph-c} shows the nearest neighbor graph for \{{\tt
Home,Photo}\}, and illustrates another type of inconsistency. In this
case, there is no majority consensus among the neighbors of \{{\tt
Home,Photo}\}. A similar situation exists in the graph for \{{\tt
Home,Document}\}, which we do not show due to space constraints. If we
look at the two graphs in Figure~\ref{fig:nn-graph-a} and
Figure~\ref{fig:nn-graph-c}, we realize that changing the policy
decision of \{{\tt Home,Memo}\} removes both the inconsistencies. Thus,
\pbe capitalizes on the possibility that a few examples may cause the
most noise, and recommends the user to change their labels.  In our
algorithm described in Section~\ref{sec:active-learning}, we describe
graph invariants to identify noise, and a greedy algorithm to find the
optimal change. Section~\ref{sec:results} demonstrates that our
interactive approach finds {\em five times} as many errors as manual
reviews by users.

Note that we do not claim to detect all errors, as the users' examples
may be completely consistent, but may still have errors. Instead, we
recommend a best effort approach for engaging the user in detecting
potential errors. 

\vspace{-1em}
\section{The \pbes Algorithm}
\label{sec:algorithm}
\vspace{-0.5em}

This section describes our algorithm for predicting policy decisions,
and the active learning approach. As stated previously, distance between
policy examples is the distance between their scenarios, and policy
decisions are the labels for the scenarios.

Our policy scenarios are Boolean functions over $n$ variables (i.e.,
tags), denoted by ${\cal B}_n$. However, we restrict our attention to
functions that are conjunctions of variables (e.g. $x_1 \wedge x_3
\wedge x_5$). Such a function $f$ can be represented as a set $I(f)
\subseteq \{ 1,2,\cdots,n \}$ (e.g., if $f = x_1 \wedge x_3 \wedge x_5$,
then $I(f) = \{ 1,3,5 \}$). Our policy scenarios belong to this
restricted class (denoted by ${\cal P}_n$). 

We had two requirements for the learning-algorithm to infer policy
decisions: {\sf (I):} non-parametric (does not rely on models with a
fixed set of parameters).  {\sf (II):} easy explanation (easy to present
to the user how the policy was inferred). For this reason we chose a
variant of the {\it $k$ nearest neighbor (kNN)} classifier~\cite{mur12}.
A kNN algorithm simply ``looks at'' the $k$ points in the training set
that are nearest to the test input $x$, counts how many members of each
class are in the set, and returns that empirical fraction as the
estimate.

Recall that our goal is to label a policy scenario $p \in {\cal P}_n$
with the decision $1$ (i.e., {\sf allow}) or $0$ (i.e., {\sf deny}). We
are also given a set of policy scenarios along with known labels (i.e.,
policy decisions).
Our algorithm is inspired by the kNN
algorithm and works as follows: given a new policy scenario $p \in {\cal
P}_n$ with an unknown label, we find the set of $k$ policy scenarios
$N(p) \; = \; \{ p_1,\cdots,p_k \}$ {\em closest} to $p$ according to
the metric $\mu$ (described in the next subsection) and then associate
the label to $p$ that corresponds to the majority labels of the policy
scenarios in $N(p)$. Our variant of kNN only considers scenarios at the
closest distance for inclusion in $N(p)$.  We describe how to address
situations with no majority in Section~\ref{sec:no-majority}.

We use active learning to assist the user in correcting potential
labeling errors in the user's policy examples. When we find that certain
conditions are not true (e.g., the label of a policy scenario $q \in
{\cal P}_n$ is different from the majority label among its neighbors
$N(q)$), we recommend a change in the label~(e.g.,~change {\sf allow} to
{\sf deny}).  We now describe our metric $\mu$, its weighted form
$\mu_w$, and the active learning phase. We design a new metric as
integrating weights into existing metrics (e.g., jaccard distance) may
incur significant re-engineering.

\vspace{-1em}
\subsection{The Metric}
\label{sec:metric}
\vspace{-0.5em}

Let $f$ and $g$ be two Boolean functions over $n$ variables
$x_1,x_2,\cdots,x_n$. A metric between $f$ and $g$ (denoted by
$\mu(f,g)$) can be defined as follows:\\
\vspace{-1em}
\[
1 - \frac{\sharp ( f \oplus g )}{2^n}
\vspace{-0.5em}
\]
Where $\oplus$ represents exclusive-or and $\sharp (h)$ is the number of
satisfying assignments of the Boolean function $h$. Recall that computing
the number of satisfying assignments of a Boolean function is a hard
problem ($\sharp$-P complete~\cite{ab09}). However, for our special case
where scenarios are conjunctions of variables, this metric is easy to
compute. Next, we describe the metric for the functions in the set
${\cal P}_n$. 

Consider two functions $f_1$ and $f_2$.  Let $[n] = \{1,2,\cdots,n\}$.
Consider three sets of indices $I_{1,2}$ (variables neither in
$f_1$ nor $f_2$), $I^1_2$ (variables in $f_1$ but not in
$f_2$) and $I^2_1$ (variables in $f_2$ but not in $f_1$);
i.e.,  $I_{1,2} = [n] \setminus (I(f_1) \cup I(f_2))$, $I^1_2 = I(f_1)
\setminus I(f_2)$, and $I^2_1 = I(f_2) \setminus I(f_1)$.  An assignment
$\sigma$ is a Boolean vector of size $n$ of the form $\langle
b_1,b_2,\cdots,b_n \rangle$ and $f(\sigma)$ denotes the value of the
function $f$ for assignment $\sigma$.  

Consider an assignment $\sigma \; = \; \langle b_1,b_2,\cdots,b_n
\rangle$ such that $f_1(\sigma) = 1$ and $f_2 (\sigma) = 0$. Then for
all $i \in I(f_1)$, $b_i = 1$, and there should be at least one $i \in
I^2_1$ such that $b_i = 0$. For $i \in I_{1,2}$, $b_i$ can assume any
value. Consider all the indices in $I^2_1$. There should be at least one
$j \in I^2_1$, such that $b_j = 0$ (we want $f_2 (\sigma) = 0$).
Therefore, the number of satisfying assignments $\sigma$, such
that $f_1(\sigma) = 1$ and $f_2 (\sigma) = 0$ is
\vspace{-1em}
\[
2^{k_{1,2}} (2^{k_2} - 1)
\vspace{-0.5em}
\]
where $k_{1,2} = |I_{1,2}|$ and $k_2 = |I^2_1|$.
Explanation for the formula is as follows: all variables with indices
in the set $I_{1,2}$ can be given any value (resulting 
in the term $2^{k_{1,2}}$). All the variables with
indices in $I^2_1$ can be given any values as long as one of them 
is $0$, so  an assignment where all variables with indices in $I^2_1$
is assigned $1$ is excluded (this results in the term $2^{k_2}-1$).

A symmetric argument shows that the number of satisfying assignments
$\sigma$ such that $f_1 (\sigma)=0$ and $f_2 (\sigma)=1$ is 
\vspace{-1em}
\[
2^{k_{1,2}} (2^{k_1} - 1)
\vspace{-0.5em}
\]
where $k_{1,2} = |I_{1,2}|$ and $k_1 = |I^1_2|$.
Adding the two terms, we have that $\sharp (f_1 \oplus f_2)$ is
$2^{k_{1,2}} (2^{k_1}+2^{k_2}-2)$. Therefore, the metric $\mu
(f_1,f_2)$ in this case is:
\[
1 - \frac{2^{k_{1,2}} (2^{k_1}+2^{k_2}-2)}{2^n}
\vspace{-0.5em}
\]
where $k_{1,2} = |I_{1,2}|$, $k_1 = |I^1_2|$, and $k_2 =
|I^2_1|$. Intuitively, $k_1$ is the number of variables that
appear in $f_1$ but not in $f_2$, $k_2$ is the number of variables that
appear in $f_2$ but not in $f_1$, and $k_{1,2}$ is the number of
variables that appear in neither $f_1$ or $f_2$.  Let $k = n - k_{1,2}$,
which is the number of variables that appear in $f_1$ and $f_2$ (i.e.,
$k = |I(f_1) \cup I(f_2)|$). The metric $\mu (f_1,f_2)$  can 
be simplified as follows:
\vspace{-0.5em}
\[
\mu (f_1,f_2) = 1 - \frac{2^{k_1}+2^{k_2}-2}{2^k}
\vspace{-0.5em}
\]

Note that higher values of $\mu$ indicate closeness.  

\vspace{-1em}
\subsection{The Weighted Metric} 
\label{sec:weighted-metric}
\vspace{-0.5em}

For security policies, some variables are more important than others;
e.g., recall the {\tt Home} tag from Bob's policy in
Section~\ref{sec:overview}. To incorporate the importance of variables
we introduce a weighted version of our metric.
As before, we will
consider a Boolean function over $n$ variables $x_1,x_2,\cdots,x_n$.
However, in this case we have two weights $w_i^0$ and $w_i^1$ associated
with each index $1 \leq i \leq n$. The weight associated with an
assignment $\sigma=\langle b_1,\cdots,b_n \rangle$ (denoted as
$w(\sigma)$) is
\vspace{-0.5em}
\[
\sum_{i=1}^n w_i^0 (1-b_i) + w_i^1 b_i \;.
\vspace{-0.5em}
\]
Given a set of Boolean assignments $S$, define $w(S)$ as 
$\sum_{\alpha \in S} w(\alpha)$ -- the sum of weights of
all assignments in $S$.  Given a Boolean function
$f$, $w(f)$ is the weight of the set of satisfying assignments 
of $f$. Using a simple 
recursive argument, the weight of all 
$2^n$ assignments $\{ 0,1 \}^n$ is:
\vspace{-0.5em}
\[
\prod_{i=1}^n (w_i^0+w_i^1)
\vspace{-0.5em}
\]

Given $n$ pair of weights $(w_1^0,w_1^1),\cdots,(w_n^0,w_n^1)$, a weighted metric between 
two Boolean functions $f$ and $g$ (denoted as $\mu_w(f,g)$) is defined as follows:
\vspace{-0.5em}
\[
1-\frac{w(f \oplus g)}{\prod_{i=1}^n (w_i^0+w_i^1)}
\vspace{-0.5em}
\]
Note that if for all $i$ we have $w_i^0=w_i^1=1$, we get the previous metric (i.e.,
the unweighted case).

As before, consider two Boolean functions $f_1$ and $f_2$ with index
sets $I(f_1)$ and $I(f_2)$. Let the index sets $I_2^1$ and $I_1^2$
be as defined before. Define the following three quantities:
\vspace{-0.5em}
\begin{eqnarray*}
z_1 & = & \prod_{i \in I_2^1} (w_i^0+w_i^1) \; - \;  \prod_{i \in I_2^1} w_i^0 \\
z_2 & = & \prod_{i \in I_1^2} (w_i^0+w_i^1) \; - \;  \prod_{i \in I_1^2} w_i^0 \\
z & = & \prod_{i \in I(f_1) \cup I(f_2) } (w_i^0+w_i^1) 
\end{eqnarray*}

The metric $\mu_w (f_1,f_2)$ can be defined as:
\vspace{-1em}
\[
\mu_w (f_1,f_2) = 1-\frac{z_1+z_2}{z}
\vspace{-0.5em}
\]
The argument is exactly same as before. The reader can check that for the unweighted
case (i.e. for all $i$ we have $w_i^0=w_i^1=1$) we get the previous metric back. 

\myparagraph{Setting weights} Next we describe an algorithm to set
weights. Given a set of variables $V \; = \; \{x_1,\cdots,x_n\}$,
suppose we are given a partial order $\preceq$ on $V$ (e.g., $x_i
\preceq x_j$ means that $x_j$ is more ``important'' than $x_i$). Next we
construct a function $L: V \rightarrow [ n ]$ that assigns integers
between $1$ and $n$ to each variable in $V$ and has the property that
$x_i \preceq x_j$ and $j \not= i$ implies that $L(x_i) >
L(x_j)$.\footnote{ Such a function can be constructed by topologically
sorting a directed graph whose nodes are $V$ and there is an edge from
$x_j$ to $x_i$ ($j \not= i$) iff $x_i \preceq x_j$.} We can assign
higher weights $w_i^1$ to variables that have a lower value according to
the function $L$ and set all the weights $w_i^0$ to $1$. 
Note that it is not necessary to precisely define a mechanism for
assigning weights, as long as the ordering imposed by $L$ is preserved.

\vspace{-1em}
\subsection{Active Learning}
\label{sec:active-learning}
\vspace{-0.5em}

Ideally, users would provide accurate examples to \pbe.  However, as
even expert users are not always accurate~\cite{yho+14,irc15}, we expect
a small margin of error in the policy decisions provided by the user;
e.g., a typo resulting in 1 being accidentally marked as 0. We use
active learning to find and correct potentially incorrect policy
decisions, by asking users to relabel certain chosen scenarios.
Relabeling samples to remove errors has been shown to be effective even
with non-experts by prior work~\cite{spi08}.  


In our approach, the scenarios and their nearest neighbors are arranged
as a graph, which allows us to relabel existing scenarios in a
systematic manner if certain invariants on the graph are not true. In
other words, the graph we are about to describe gives us a systematic
way to evaluate the conditions that may indicate user error.

Let $G = (V,E,L_V,L_E)$ be a $4$-tuple where $V \subseteq {\cal P}_n$ is
the set of labeled policy scenarios, $E\subseteq V \times V$ is the set
of edges, $L_V$ maps each vertex $v \in V$ with a label $1$ (signifying
{\sf allow}) and $0$ (signifying {\sf deny}), and $L_E$ labels each edge
$e \in E$ with a non-negative real value (i.e., $L_E (v,v')$ is $\mu
(v,v')$, which is the distance between the scenarios $v$ and $v'$).  The
set of neighbors $N(v)$ of a vertex $v \in V$ is the set $\{ v' \; | \;
(v,v') \in E \}$ and intuitively represents all the nearest-neighbors of
the policy scenario $v$.

\noindent
{\bf (Inv-1): Majority label exists.} This invariant states that for all $v \in V$, 
its set of neighbors  $N(v)$ have a majority label 
(i.e,. more than $\frac{|N(v)|}{2}$ vertices in $N(v)$ have the same label
$L_V (v)$).

\noindent
{\bf (Inv-2): Agreement with the majority label.} This invariant states that
if invariant Inv-1 is true, then for every $v \in V$ its label
$L_V (v)$ agrees with the majority label of its neighbors $N(v)$.

Intuitively we want the graph $G$ corresponding to our policies to
satisfy invariants Inv-1 and Inv-2. If the graph $G$ violates either of
the invariants, then we recommend relabeling of the policy scenarios to
the user.

\begin{figure}[t]
  \centering
  \captionsetup[subfigure]{font=small,labelfont={bf,sf},width=1.1in}
  \subfloat[][$p$ violates Inv-1, i.e., no majority
  label]{\includegraphics[width=0.6in]{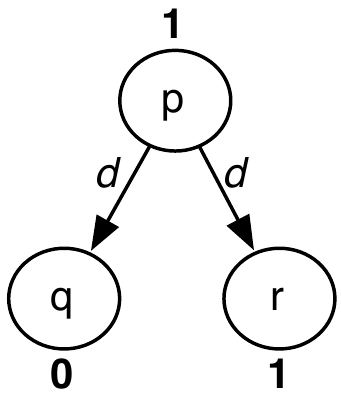}\label{fig:invariant1}}\qquad\qquad\qquad
  \subfloat[][$p$ violates Inv-2, i.e., disagrees with majority
  label]{\includegraphics[width=0.6in]{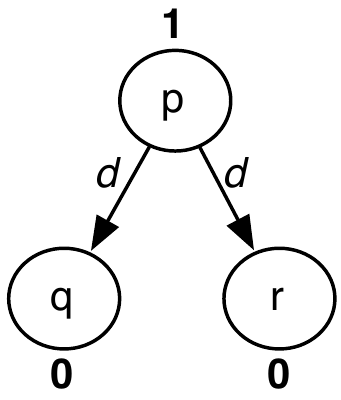}\label{fig:invariant2}}
  \vspace{-0.5em}
  \caption{NN graph for vertex $p$, consisting of neighbors $q$ and $r$ at the closest distance $d$, violates Inv-1 and Inv-2.}
  \label{fig:invariants}
  \vspace{-1.5em}
\end{figure}

Figure~\ref{fig:invariants} shows instances of the graph for some vertex
$p$ that violate the invariants. In Figure~\ref{fig:invariant1}, there
is no majority label among $p's$ neighbors, which can be resolved by
relabeling either $q$ or $r$.  Further, in Figure~\ref{fig:invariant2},
the label on $p$ disagrees with the majority, which can be resolved by
relabeling $p$.   

We use a simple greedy approach to recommend changes:
Consider a function ${\cal V}(G)$ that counts the total violations of
both Inv-1 and Inv-2 in graph $G$. Further, consider a function ${\cal
C}(v,G)$ that measures the impact of a potential label change on
violations, i.e., returns the decrease in ${\cal V}(G)$ after a
temporary change in the label of $v$ (i.e., $L_V (v)$).  The
label change that causes the maximum decrease in ${\cal V}(G)$ is
optimal.  Therefore, at each iteration, we find the optimal vertex,
$v_{opt}$, by maximizing ${\cal C}(v,G)$ over all $v\in~V\setminus
V_{visited}$, where $V_{visited}$ is the set of all vertices that have
been recommended to the user previously. We add $v_{opt}$ to
$V_{visited}$, and recommend the user to change $L_V (v_{opt})$.  If the
user accepts, we change $L_V (v_{opt})$.  We reiterate until all the
vertices are visited or until there are no more violations.

\vspace{-1em}
\subsection{Prediction with No Majority}
\vspace{-0.5em}
\label{sec:no-majority}
As described previously, we predict the label (i.e., policy decision)
for a new policy scenario $p$ as the majority label of its nearest
neighbors $N(p)$. If there is no majority label, we use the following
method for prediction:

We eliminate the first neighbor that is not a {\em mutual} neighbor,
i.e., if there is a labeled policy scenario $q$ such that $q \in N(p)$
but $p \notin N(q)$, we remove $q$ from $N(p)$, thereby converging on a
majority.  In case such elimination is not possible, i.e., if all
neighbors in $N(p)$ are mutual neighbors, we deny by default.  Our
method considers the value of the distance between neighbors to resolve
a tie, instead of randomly discarding one scenario (i.e., by considering
only an odd number of scenarios in $N(p)$). In
Section~\ref{sec:results}, we demonstrate that \pbe performs better than
a baseline of random guessing, and that such cases were rare, i.e.,  
less than 6\% test scenarios had no majority, and less than 3\% were
denied by default.

\vspace{-1em}
\section{Evaluation}
\label{sec:eval}
\vspace{-0.5em}

We performed an IRB-approved feasibility study with expert users to
evaluate the effectiveness of our approach.  We chose experts under the
hypothesis that they prefer more complex policies, the complexity of
which makes them challenging to predict.  Support for this hypothesis
comes from the fact that non-expert users are more likely to employ
binary security practices, (e.g., only visiting known sites rather than
deciding based on security-related attributes like usage of
https~\cite{irc15}) and evidence that knowledge of security risks can
increase sensitivity to security when making data
decisions~\cite{rwb12}.  Note that no personally identifiable
information (PII) was collected from the participants.  

We plan to release our tool and source code after publication, to allow
a broader audience to use \pbe. A sanitized version of our dataset will
also be released.
The following research questions motivate our study:\vspace{-0.5em}
\begin{enumerate}[label=\textbf{RQ\arabic*},ref=\textbf{RQ\arabic*}]\setlength{\itemsep}{-0.3em} 
  \item \label{rq:accuracy} {\it How accurate are our predictions for
    random, unlabeled scenarios that may occur at runtime?}
  \item \label{rq:accuracy-causes} {\it What are the causes for incorrect
    predictions?}
  \item \label{rq:user-error} {\it Do users make mistakes in their
    examples?} 
  \item \label{rq:user-error-auto} {\it Does our active learning
    approach help the user find mistakes in their examples?}
\end{enumerate}
\vspace{-0.5em}
This section describes the study setup, the data collection and
experiments. Section~\ref{sec:results} describes the results.  Due to
space constraints, this section describes the core methodology of the
feasibility study; the literal scripts used during
the study can be found in Appendix~\ref{app:feasibility_details}.

\vspace{-1em}
\subsection{Study Setup}
\label{sec:study_setup} 
\vspace{-0.5em}

In this study, participants were asked to consider a smartphone
environment, where personal and work data would be at risk of
unauthorized exfiltration from the device.  We now describe the
participants (i.e., expert users), policy targets and information tags
involved in the study.


\myparagraphs{Expert Users} We recruited 8 graduate
student researchers from a security research lab for this study (denoted
as \user{1}$\rightarrow$\user{8}). Our participants had at least 1
academic year of experience (2.5 years on average) in security research
at the time of this study, including at least one research project and
two graduate-level courses in security or privacy.  
We use the security-focused course-work and research as an indicator of
general security-awareness, and assume the participants to be well-aware
of their own security and privacy requirements.  Additionally, we
confirmed that all our participants used their smartphones for both work
and personal data.  Finally, through an informal discussion of
participant background knowledge, we confirmed that the participants
were aware of the threat of exfiltration of work and personal data by
third party applications on smartphones, as discussed by prior work
(e.g., TaintDroid~\cite{egc+10}).

\begin{table}[t]
\centering
\scriptsize
\caption{Policy targets and the actions they control.
}
\vspace{-1em}
\label{tbl:policy}
\begin{tabular}{r|L}
  \Xhline{2\arrayrulewidth}
 {\bf Policy Target}	& {\bf Action Controlled}\\ 
  \Xhline{2\arrayrulewidth}
  {\em WorkCloud} & Export of data {\em to} the enterprise cloud\\\hline
  {\em PersonalCloud} & Export of data {\em to} the personal
  cloud\\\hline
  {\em WorkEmailApp} & Export of data {\em by} the enterprise email
  app\\\hline
  {\em PersonalEmailApp} & Export of data {\em by} the personal email
  app\\\hline
  {\em SocialApp} & Export of data {\em by} the social network
  app\\
  \Xhline{2\arrayrulewidth}
  \end{tabular} 
  \vspace{-1.5em}
\end{table}
 
\myparagraphs{Policy Targets}
Table~\ref{tbl:policy} provides the policy targets (i.e., policies) used
in our study. The targets are similar to the {\em WorkCloud} target
discussed in Section~\ref{sec:pbe}, and either {\sf (a)} restrict the
destination Web domain to which data can be exported (i.e., {\em
WorkCloud} and {\em PersonalCloud}) or {\sf (b)} restrict the exporters
(i.e., applications) that are permitted to export data; i.e., {\em
WorkEmailApp}, {\em PersonalEmailApp} and {\em SocialApp} regulate
export by the user's work email client, personal email client, and
social network client (e.g., the Facebook app) respectively.  As
described previously, each target is treated as an independent policy. 

\myparagraphs{Information Tags} We provided users with 9 predefined
secrecy tags, based on tags available in popular note-taking
applications (e.g., Google Keep, Evernote). To enable our experts to
create any complex policy they desired, we allowed them to create new
tags as well. The tags (user-created or predefined) were primarily of
two kinds, namely tags that defined the location or time at which the
information was created (e.g., {\tt Work}, {\tt Afterhours}) or the
type or class of information (e.g., {\tt Receipt}, {\tt
WhiteboardSnapshot}).  The tags used in this study are provided in
Appendix~\ref{app:tags}.

\vspace{-1em}
\subsection{Data Collection}
\label{sec:user_study}
\vspace{-0.5em}

This section describes the approach used for collecting the policy
examples and weights from participants.

\myparagraphs{1. Collecting Policy Examples} Participants were provided
with our predefined tags, but were also allowed to create their own
tags. Participants were instructed that they could combine tags into
complex scenarios for creating examples.  We placed no constraint on the
number of example scenarios each participant could provide. For each
scenario, participants were required to label policy decisions for the 5
targets described previously in Table~\ref{tbl:policy}.  Note
that we collected labels for two more targets, but discarded them
before testing to reduce user fatigue.  

A preliminary analysis of the examples collected from our participants
led to two interesting observations: {\sf (1)} Our participants created
a total of 31 unique tags, out of which about 58\% (or 23) were specific
to individual participants, while only 7 tags were commonly used by all
in their examples, and {\sf (2)} Out of the 246 example scenarios
collected across participants, over 76\% were specific to individual
participants, and only 7 were common among all 8 participants. These
observations indicate that relevant {\em data-use scenarios may be
unique to the individual, even among student researchers from the same
research lab}, further motivating our research into generating
user-specific policies for user-specific data.

\myparagraphs{2. Obtaining Weights} On average, each participant used
about 14 unique tags in their examples. As ordering a large number of
tags can be tiring, we categorized  tags into semantic groups. The
participants were provided with this semantic grouping, and were first
{\sf (1)} allowed to customize group memberships of tags as per their
understanding, and then {\sf (2)} instructed to provide a partial order
over the groups in a spreadsheet.  Participants were provided with a
basic partial order generated by the authors, and could start from
scratch, or customize the provided ordering.  We confirmed each partial
order relation by reading it out to the user; e.g., by asking if ``j is
more important than i'' to confirm {\tt i;j}. We then transformed the
orders to weights using the approach described in
Section~\ref{sec:weighted-metric}. For additional illustration,
Appendix~\ref{app:weights_script} provides \user{1}'s tag groups
(Figure~\ref{fig:tag_groups}) partial order on the groups (
Figure~\ref{fig:p1_order}).

Finally, participants were informed that they could provide different
partial orders for different policies, but most participants chose to
keep a single general order.  We describe the impact of
this decision in Section~\ref{sec:analysis_of_results}.  

\vspace{-1em}
\subsection{Experiments}
\label{sec:experiments}
\vspace{-0.5em}
This section describes the experiments for identifying user errors
and testing prediction for random scenarios.

\myparagraphs{1. Identifying Errors}
\begin{lstlisting}[basicstyle=\ttfamily\scriptsize,float=tp,caption={A
  suggestion made 
  by the \pbes algorithm during the interactive review
  process.},belowcaptionskip=-6mm,label=lst:suggestion,emph={Suggestion},emphstyle=\bfseries]
Suggestion: For {Note}, WorkCloud = DENY. Agree?(y/n)
\end{lstlisting}
The review of examples was carried out 3 months after the initial
specification, as most participants were unavailable over the summer
break.  
We performed a two-step experiment to help participants identify and
correct errors in their policy decisions.  

First, participants performed a manual review of their initial
specification. Participants were provided with a spreadsheet containing
their policy examples (one sheet per policy target), and could change
any policy decision they desired.  For each update, participants were
instructed to indicate a cause to justify the change (e.g., correcting
an error, change of mind, inability to decide). Finally, participants
provided a justification for each change (e.g., ``Work is
confidential''), providing the helpful context for analyzing the results
(Section~\ref{sec:analysis_of_results}).

After the manual review, we performed a \pbe-assisted review using the
approach described in Section~\ref{sec:active-learning}. We treat each
participant-policy combination as a separate policy specification
problem; hence, a separate review was performed for each such case
(i.e., 8 users and 5 policies make 40 total cases). As we used the
changed examples from the manual review; any errors discovered using
this approach were additional. Our algorithm presented the participant
with a series of suggestions (i.e., examples with corrected policy
decisions, as shown in Listing~\ref{lst:suggestion}).  If the
participant accepted the suggestion, we confirmed with the participant
that the original decision was in error, and recorded it as an error
found by \pbe.  If the participant rejected, we asked for a short
justification to understand the participant's policy preferences. We
stopped at 15 suggestions for each participant-policy case to limit
fatigue.  

\myparagraphs{2. Testing with Random Scenarios}
For each participant, we randomly generated $n/2$ new policy scenarios,
where $n$ was the number of scenarios initially provided by the
participant.  The random scenarios were created
with the tags used in the participant's initial examples.  The intuition
is that the tags provided by the participant are relevant to the
participant; hence scenarios composed of them must be relevant as
well. To mitigate labeling fatigue, the random scenarios included at
most 3 tags. 

Participants provided the ground truth policy decisions for their
test scenarios, for each of the five policy targets.  Apart from
indicating ``Allow'' or ``Deny'', participants were also provided the
``I don't know'', in which case we substituted the scenario with another
random test scenario.  We predicted the policy decision for each test
scenario using our algorithm. We then asked participants to confirm
their decisions for incorrect predictions, provide short justifications,
and conducted short, informal interviews that helped us gain insight
into the decisions.

\vspace{-1.5em}
\section{Results}
\label{sec:results}
\vspace{-0.5em}

This section describes the results of our experiments, i.e., \pbe's
accuracy in predicting policy decisions for new scenarios, and its
effectiveness in assisting participants in finding incorrect policy
decisions in their examples.  We start by briefly describing the
datasets collected during the initial policy specification and testing;
a detailed split across participants can be found in
Appendix~\ref{app:datasets}.

\myparagraphs{Specification dataset} 
The 8 participants provided 246 example scenarios in total, with policy
decisions for 5 policy targets, resulting in a total of 1,230 initial
labeled policy examples.  

\myparagraphs{Testing dataset}
We generated a total of 122 random test scenarios across 8 participants,
which when labeled with ground-truth policy decisions by participants
for 5 policies, resulted in 610 test examples. 

\vspace{-1em}
\subsection{Accuracy of Predictions}
\vspace{-0.5em}
\begin{table*}[t]
\scriptsize
\centering
\caption{Accuracy of \pbes in comparison with the {\em CoinFlip}
(abbreviated to CF) baseline, for all 40 user-policy
cases. Cases where the accuracy of CF is greater are highlighted
in {\bf bold}.}
\vspace{-1em}
\label{tbl:coinflip_perf}
\setlength{\tabcolsep}{5pt}
\begin{tabular}{c|cr|cr|cr|cr|cr|cr|cr|cr}
  \Xhline{2\arrayrulewidth}
  ~		    & \multicolumn{2}{c|}{\bf \user{1}}	& \multicolumn{2}{c|}{\bf \user{2}}	& \multicolumn{2}{c|}{\bf \user{3}} & \multicolumn{2}{c|}{\bf \user{4}} & \multicolumn{2}{c|}{\bf \user{5}} & \multicolumn{2}{c|}{\bf \user{6}} & \multicolumn{2}{c|}{\bf \user{7}} & \multicolumn{2}{c}{\bf \user{8}}\\ 
{\bf Policy Target}		&{\bf \pbe}&{\bf CF}&{\bf \pbe}&{\bf CF}&{\bf \pbe}&{\bf CF}&{\bf \pbe}&{\bf CF}&{\bf \pbe}&{\bf CF}&{\bf \pbe}&{\bf CF}&{\bf \pbe}&{\bf CF}&{\bf \pbe}&{\bf CF} \\
  \Xhline{2\arrayrulewidth}
    WorkCloud	    &	0.96 & 0.50	& 0.73 & 0.48	& 0.66 & 0.49	& 0.83 & 0.49	& 0.56 & 0.51	& 0.93 & 0.50	& 0.90 & 0.49	& 0.67 & 0.49\\
    PersonalCloud   &	0.77 & 0.50	& 0.55 & 0.5	& 1.00 & 0.50	& 0.75 & 0.52	& 0.75 & 0.51	& {\bf 0.50} & {\bf 0.51}   & {\bf 0.40} & {\bf 0.51}   & 0.71 & 0.49\\
    WorkEmailApp    &	0.96 & 0.50	& 0.55 & 0.51	& 0.83 & 0.47	& 0.83 & 0.47	& 0.63 & 0.51	& 0.86 & 0.51	& 0.90 & 0.50	& 0.76 & 0.49\\
    PersonalEmailApp &  0.77 & 0.51	& 0.55 & 0.49	& 1.00 & 0.50	& 0.75 & 0.48	& 0.63 & 0.51	& {\bf 0.50} & {\bf 0.51}	& 0.50 & 0.48	& 0.67 & 0.51\\
    SocialApp	     &  0.81 & 0.51	& 0.73 & 0.49	& 0.75 & 0.52	& 0.92 & 0.49	& 0.94 & 0.50	& 1.00 & 0.51	& 1.00 & 0.52	& 0.91 & 0.50\\
  \Xhline{2\arrayrulewidth}
    Average & 0.85 & 0.50   &   0.62 & 0.49  &      0.85 & 0.50   &    0.82 & 0.49   &    0.70 & 0.51   &   0.76 & 0.51   &   0.74 & 0.50   &   0.74 & 0.50\\  
    \Xhline{2\arrayrulewidth}
  \end{tabular} 
\vspace{-0.5em}
\vskip1pt
\begin{flushleft}
{\footnotesize $^{*}$The {\em CoinFlip} baseline values shown are the
mean of 50 executions with a 95\% confidence interval less than 0.03.}
\end{flushleft}
\vspace{-2em}
\end{table*}

\begin{table*}[t]
\scriptsize
\centering
\caption{Accuracy of \pbes in comparison with the {\em MostFreq}
(abbreviated to MF) approach, for all 40 user-policy
cases. Cases where the accuracy of MF is greater are highlighted
in {\bf bold}.}
\vspace{-1em}
\label{tbl:mostfreq_perf}
\setlength{\tabcolsep}{5pt}
\begin{tabular}{c|cr|cr|cr|cr|cr|cr|cr|cr}
  \Xhline{2\arrayrulewidth}
  ~		    & \multicolumn{2}{c|}{\bf \user{1}}	& \multicolumn{2}{c|}{\bf \user{2}}	& \multicolumn{2}{c|}{\bf \user{3}} & \multicolumn{2}{c|}{\bf \user{4}} & \multicolumn{2}{c|}{\bf \user{5}} & \multicolumn{2}{c|}{\bf \user{6}} & \multicolumn{2}{c|}{\bf \user{7}} & \multicolumn{2}{c}{\bf \user{8}}\\ 
{\bf Policy Target}		& {\bf \pbe} & {\bf MF} & {\bf \pbe}  & {\bf MF} & {\bf \pbe}  & {\bf MF} & {\bf \pbe}  & {\bf MF} & {\bf \pbe}  & {\bf MF} & {\bf \pbe}  & {\bf MF} & {\bf \pbe}	& {\bf MF} & {\bf \pbe}  & {\bf MF} \\
  \Xhline{2\arrayrulewidth}
    WorkCloud	    &	0.96 & 0.85	& 0.73 & 0.63	& 0.66 & 0.17	& 0.83 & 0.75	& {\bf 0.56} & {\bf 0.88}	& 0.93 & 0.71	& 0.90 & 0.70	& 0.67 & 0.43\\
    PersonalCloud   &	0.77 & 0.69	& {\bf 0.55} & {\bf 0.91}	& 1.00 & 1.00	& 0.75 & 0.75	& 0.75 & 0.75	& 0.50 & 0.21   & 0.40 & 0.20   & 0.71 & 0.71\\
    WorkEmailApp    &	0.96 & 0.85	& {\bf 0.55} & {\bf 0.64}	& 0.83 & 0.17	& 0.83 & 0.75	& {\bf 0.63} & {\bf 0.81}	& 0.86 & 0.71	& 0.90 & 0.80	& 0.76 & 0.48\\
    PersonalEmailApp &  0.77 & 0.69	& {\bf 0.55} & {\bf 0.91}	& 1.00 & 1.00	& 0.75 & 0.75	& {\bf 0.63} & {\bf 0.81} & {\bf 0.50} & {\bf 0.79} & {\bf 0.50} & {\bf 0.80}	& {\bf 0.67} & {\bf 0.71}\\
    SocialApp	     &  0.81 & 0.73	& 0.73 & 0.46	& {\bf 0.75} & {\bf 0.92}	& 0.92 & 0.58	& {\bf 0.94} & {\bf 1.00} & 1.00 & 1.00	& 1.00 & 1.00	& 0.91 & 0.91\\
  \Xhline{2\arrayrulewidth}
    Average & 0.85 & 0.76   &   {\bf 0.62} & {\bf 0.71}  &	0.85 & 0.65   &    0.82 & 0.72	 &    {\bf 0.70} & {\bf 0.85}   &   0.76 & 0.68   &   0.74 & 0.70   &	0.74 & 0.65\\
  \Xhline{2\arrayrulewidth}
  \end{tabular} 
\vspace{-0.5em}
\end{table*}
 
Our algorithm predicted decisions for all of the participants' test
scenarios.\footnote{We test with random
samples instead of cross-validation, as the latter is generally
used to test ``models'', i.e., in supervised learning.} The actual
prediction time was negligible (i.e., less than 1 second for all the
examples per participant).  Further, for less than 6\% (36 out of 610)
of our test examples we had no majority label (i.e., a tie).  Applying
the tiebreaker discussed in Section~\ref{sec:no-majority} resolved 19
of these ties, while the rest (i.e., 3\% or 17 out of 610) were denied
by default. We now discuss the accuracy of \pbe's predictions.

On comparing our predicted decisions with ground-truth decisions
provided by participants, we observe that \pbes predicts policy
decisions with an average accuracy of over 76\% across all participants
(\ref{rq:accuracy}).  When analyzing the accuracy, it is important to
note that each participant-policy combination is treated as an
independent policy specification problem, and hence forms a separate
test case. We first define a baseline and naive approach against which
we evaluate \pbe's accuracy. 

\myparagraphs{1. The CoinFlip baseline} The {\em CoinFlip} baseline
provides the measure of accuracy of random guessing, with an equal
probability of a 0/1 outcome on each flip. 

\myparagraphs{2. The MostFreq naive approach} We define {\em MostFreq} as
an approach that predicts the most frequent or majority policy decision
from the specification dataset, {\em independently for each
participant-policy problem}.  For example, if \user{1} generally allows
export to {\em WorkCloud}, {\em MostFreq} will predict allow for all new
test examples for that the \user{1}-{\em WorkCloud} policy specification
problem.  The insight behind {\em MostFreq} is that a naive learner is
likely to pick the majority class to benefit from the consistent trend
in the participant's policy decisions.

Table~\ref{tbl:coinflip_perf} shows the comparison of \pbe's accuracy
with {\em CoinFlip}, for each of the 40 participant-policy cases. \pbe
not only performs better in terms of average accuracy (i.e., 76$>$50),
but also for most (i.e., all but 3, or 92\%) of the participant-policy
problems.

Table~\ref{tbl:mostfreq_perf} shows a comparison between the performance
of \pbe and the naive approach {\em MostFreq}.  \pbes not only performs
better than {\em MostFreq} in terms of average accuracy (i.e., 76$>$71),
but also in 29 out of 40 (i.e., 72.5\%) participant-policy cases, and
for 75\% of the participants. Note that although {\em MostFreq's}
average accuracy can be said to be close to \pbe, it has high variance,
with accuracy dropping to 17\% in some cases. This is because of {\em
MostFreq's} over-dependence on the probability distribution of the
training samples, a flaw \pbe is not susceptible to. We discuss the
causes of incorrect predictions (\ref{rq:accuracy-causes}) in
Section~\ref{sec:analysis_of_results}.

\vspace{-1em}
\subsection{Effectiveness of Active Learning}
\begin{table*}[t]
\scriptsize
\centering
\caption{The number of errors identified via manual review of examples
(abbreviated as MR), and the additional errors found in the
\pbe-assisted review. Cases where the manual review finds more errors
are highlighted in {\bf bold}.  }
\vspace{-1em}
\label{tbl:review}
 \setlength{\tabcolsep}{5pt}
\begin{tabular}{ c|cc|cc|cc|cc|cc|cc|cc|cc}
  \Xhline{2\arrayrulewidth}
  ~		    & \multicolumn{2}{c|}{\bf \user{1}}	& \multicolumn{2}{c|}{\bf \user{2}}	& \multicolumn{2}{c|}{\bf \user{3}} & \multicolumn{2}{c|}{\bf \user{4}} & \multicolumn{2}{c|}{\bf \user{5}} & \multicolumn{2}{c|}{\bf \user{6}} & \multicolumn{2}{c|}{\bf \user{7}} & \multicolumn{2}{c}{\bf \user{8}}\\ 
  {\bf Policy Target}		& {\bf MR} & {\bf \pbe} & {\bf MR} & {\bf \pbe} & {\bf MR} & {\bf \pbe} & {\bf MR} & {\bf \pbe} & {\bf MR} & {\bf \pbe} & {\bf MR} & {\bf \pbe} & {\bf MR} & {\bf \pbe} & {\bf MR} & {\bf \pbe} \\
  \Xhline{2\arrayrulewidth}
    WorkCloud	    &	0 & 0 	& 0 & 5	 & 0 & 1   & 1 & 1   & 0 & 4   & 0 & 2   & 0 & 1   & 0 & 5\\
    PersonalCloud   &	0 & 0 	& 0 & 2	 & 0 & 0   & {\bf 1} & {\bf 0}   & 0 & 3   & 2 & 3   & 0 & 0   & {\bf 4} & {\bf 2}\\
    WorkEmailApp    &	0 & 0 	& 0 & 3	 & 0 & 3   & 1 & 1   & 2 & 2   & 0 & 4   & 1 & 1   & 0 & 5\\
    PersonalEmailApp &  0 & 0 	& 0 & 3   & 1 & 2   & 0 & 1   & 0 & 6   & 0 & 5   & 0 & 0   & {\bf 2} & {\bf 1}\\
    SocialApp	     &  0 & 0 	& 0 & 4   & 0 & 3   & 1 & 1   & 0 & 1   & 0 & 5   & 0 & 0   & 0 & 0\\
  \Xhline{2\arrayrulewidth}
    Total & 0 & 0   &   0 & 17   & 1 & 9   &    4 & 4	 &    2 & 16 &   2 & 19   &   1 & 2   &	6 & 13\\
  \Xhline{2\arrayrulewidth}
  \end{tabular} 
  \vspace{-1em}
\end{table*}
 
\vspace{-0.5em}

Table~\ref{tbl:review} shows the number of labeling errors found by the
participant through the manual review, followed by the additional errors
found using \pbe's interactive approach.  Errors found by \pbe's
approach are additional as we use the corrected dataset from the manual
review for the \pbe-assisted review, as described in
Section~\ref{sec:experiments}.


Out of 1,230 total examples in the specification dataset, we observe 96
total errors (i.e., 7.8\%), with at least one labeling error in most
participant-policy cases (i.e., 30 our of 40) (\ref{rq:user-error}).
While participants identify some errors manually, \pbe's semi-automated
process helps the participant identify and correct the maximum number of
errors (80 out of 96, or about 83\%).  Further, for all 8 participants,
the total errors (across policies) found by \pbes are equal to or more
than the participant's manual review (\ref{rq:user-error-auto}).


We note that \user{1} did not find any errors manually, nor did they
agree to \pbe's recommendations as they had confidence in their
examples. Further, our accuracy is also the highest for \user{1} as seen
in Table~\ref{tbl:coinflip_perf}. However, given the absence of such a
trend in other cases, we do no claim any relation between user errors
and accuracy.


\vspace{-1em}
\section{Analysis of Results}
\label{sec:analysis_of_results}
\vspace{-0.5em}

\pbe is the first step towards our vision of a policy assistant, and
one of our objectives is to learn lessons for future work.  With this
motivation, we performed an in-depth study of our results to identify
the general causes of incorrect predictions.

We manually analyzed each of the 141 incorrectly predicted test
examples, using the following information collected during our study:
{\sf (1)} the justifications provided by the participants for their
decisions, {\sf (2)} the nearest neighbors of the test example, {\sf
(3)} the weights of the tags involved, and {\sf (4)} all examples from
the specification dataset that contain tags in common with the test
example.  The rest of this section describes the four causes of
incorrect predictions that we identified.  A detailed breakdown of the
causes across participants and policies is provided in
Table~\ref{tbl:incorrect} in Appendix~\ref{app:analysis}.

\myparagraphs{1. Misconfigured Weights} 
We found that a majority of our incorrect predictions (79 out of 141, or
over 56\%) were caused because the weights set by the participants
contradicted their actual security preferences. We confirmed our
findings using justifications from participants that clearly indicated
the tag or security preference that influenced their policy decision for
a test example. 

\begin{table}[t]
\scriptsize
\centering
\caption{A subset of policy examples specified by \user{1},
which includes only those examples that contain {\tt Work}}
\vspace{-1em}
\label{tbl:weight_cause}
\begin{tabular}{r|l|c}
  \Xhline{2\arrayrulewidth}
  {\bf No.} & {\bf Scenario}             & {\bf Policy Decision}\\
  \Xhline{2\arrayrulewidth}
  1 & \{Work, ScannedDocument\}   & deny  \\  
  2 & \{WhiteboardSnapshot, Work\} & deny\\
  3 & \{Work, BusinessCard\} & deny\\
  4 & \{Work, Audio\} & deny\\
  5 & \{Work, Postit\} & deny\\
  6 & \{Work\} & deny\\
  7 & \{Work, CalendarLink\} & deny\\
  8 & \{Work, Receipts\} & deny\\
  9 & \{Photos, Work\} & allow\\
  \Xhline{2\arrayrulewidth}
    \end{tabular} 
    \vspace{-2em}
\end{table} 

For instance, consider an incorrect prediction for \user{1}'s {\em
PersonalCloud} policy, where \pbe predicted the policy decision {\sf
allow} for the test example \{{\footnotesize \tt
WhiteboardSnapshot,Work,ScannedDocument}\}.  The user provided the
ground-truth decision of {\sf deny}, and justified with the quote ``{\em no work
data to personal cloud}''. That is, the tag {\tt Work} was confidential
and hence important to \user{1} with respect to the {\em PersonalCloud}
policy target. This preference  of {\tt Work} being important is also consistent for all but one of
\user{1}'s examples containing {\tt Work}, as shown in
Table~\ref{tbl:weight_cause}. 
However,  this importance was not reflected in the weights, i.e.,
\user{1} mistakenly assigned {\tt Work} data a lower weight (i.e.,
weight 2) by ordering it lower than personal data (i.e., weight 4). This
resulted in the test example being matched with personal examples (e.g.,
\{{\tt MedicalFacility,ScannedDocument}\}) that allowed export for
PersonalCloud.

On raising the weight of {\tt Work} to 5 (i.e., above personal tags),
the test example was correctly found closer to \{{\tt
Work,ScannedDocument}\}, resulting in a correct prediction of {\sf
deny}. Note that this increase in weight is not arbitrary, but guided by
evidence of the user's security preferences. On correcting all
misconfigured weights, we manually confirmed that our overall accuracy
rose to 89\%.  This includes most predictions for \user{2} and \user{5}
for whom \pbe had the lowest accuracy.

Since misconfigured weights caused the maximum incorrect predictions (79
out of 141, or 56\%), we investigated further, and made two interesting
observations:

\emparagraph{Observation 1: Inaccurate predictions resulted from
participants only considering privacy preferences when setting weights.}
Recall that our participants were provided with the option of setting
different weight-group orders for different policy targets in
Section~\ref{sec:user_study}. All participants (except \user{8}) set
only a general order for all 5 policy targets, which only accounted for
the their privacy preferences. As a result, higher-weighted personal
tags (e.g., {\tt MedicalFacility}, {\tt Home}) had more influence on the
policy decision, irrespective of the actual policy target. However,
participants labeled examples based on their policy target-specific
security preferences (e.g., no work to {\em PersonlCloud}. This resulted
in incorrect predictions, as seen in \user{1}'s example previously. 

Note that this phenomenon occurs only because our tags are semantically
related to the policy targets (e.g., {\tt Work} to {\em WorkCloud}).  We
confirmed that at least 26 incorrect predictions (out of 79 due to
weights) were false negatives in predicting the {\em PersonalCloud} and
{\em PersonalEmailApp} targets, because participants considers privacy
for weights, and security for labeling.

\emparagraph{Observation 2: ``Important'' may not just mean
confidential.} In at least 14 of the test-examples incorrectly predicted
due to weights, participants wanted to set a high weight for a
non-confidential tag, i.e., to declassify data if a certain tag were
present in the scenario. This was in complete contrast with the initial
understanding of the participants while setting weights, i.e., that
confidential tags would have high weights. 

\myparagraphs{2. Policy Change} 
A significant minority (30 out of 141, or over 21\%) of our incorrect
predictions resulted from a change in the participants' policies, i.e.,
when participants explicitly disagreed with an earlier assumption. 

For example,  \user{8}'s policy changed for the tag {\tt School}. During
the initial specification, \user{8} assumed {\tt School} and {\tt Work}
to be different due to off-campus employment. However, before testing,
\user{8} started working at the school, which resulted in similar
decisions for {\tt School} and {\tt Work}. \user{8} admitted to this
change during the post-testing interview. All cases in this category
were similarly confirmed.


\myparagraphs{3. Unconfirmed Policy Change} For a small number of
incorrect predictions (15 out of 141, or about 11\%), we observed a
clear contradiction between the participant's examples during
specification and testing, but could not get a confirmation from the
participant. For example, for \user{8}'s test example \{{\tt
SavedToDevice,Audio}\}, the ground truth label allows export for the
{\em WorkCloud} policy, but all except one of \user{8}'s initially
specified examples containing {\tt SavedToDevice} or {\tt Audio} deny
export to the {\em WorkCloud}. Without additional information, we
classify such contradictions as unconfirmed policy changes.


\myparagraphs{4. Tag Confusion} 
The least number of errors (i.e., 12 out of 141, or about 8.5\%) were
caused due to the ambiguity of some tags.  The location or time-based
tags (e.g., Home and Afterhours) were intended to indicate the
location or time of creation of data. However, as we did not place
strict constraints, our participants also created scenarios where such
tags could be used by themselves (e.g., the scenario \{{\tt Home}\}
could mean data created at home).  Justifications indicated that while
participants could comprehend the scenarios they had created, a few
random test scenarios (in case of 3 participants) caused confusion.  For
example, \{{\tt Afterhours,Audio,Document}\} could mean Audio created
after hours, and added to a document whose origin is unknown, or a
document created after hours, and added to an audio recording.

Finally, we exclude five incorrect predictions from \user{8}'s
test dataset from our categorization, i.e., 3 for {\em PersonalEmailApp}
and 2 for {\em WorkEmailApp}, as the participant was unable
to decide unless they knew the identity of the email receiver, which
gave us no information. We did not face this situation with any other
user or example.

\vspace{-1em}
\section{Lessons}
\label{sec:lessons}
\vspace{-1em}
The lessons we learned from our feasibility study highlight aspects of
correctly using \pbe in practice, and also motivate problems for future
work.

\myparagraphs{Lesson 1} {\em Weight assignment should reflect security},
as well as general privacy preferences. If policy targets are
semantically related to the tags, a generic weight assignment for
multiple targets may be inaccurate.


\myparagraphs{Lesson 2} {\em Addressing ``potential''
change in the policy} is imperative. The user may change some or
all of their security policy goals without informing the system.

\myparagraphs{Lesson 3} {\em The notion of importance depends on the
security goals,} as users may want to consider extremely
non-confidential data as important (i.e., declassifiers). If the
system's goal is security, then the most confidential tag may have the
highest weight; and for usability, the most non-confidential tag.

\myparagraphs{Lesson 4} {\em Tag semantics should be carefully
considered} for applying \pbes. Users may be able to reason about
examples they create, and may even desire the expressibility of 
multiple label semantics (i.e., ``created at'' or ``derived
data''); however, this expressibility may cause confusion when reasoning
about random examples.

\vspace{-1em}
\section{Future Directions}
\label{sec:future}
\vspace{-0.5em}

Our evaluation of \pbe demonstrates feasibility, and shows promise for
further exploration in this area.  We now discuss two future directions,
namely {\sf (1)} adapting \pbe for non-experts and {\sf (2)} adapting to
change.

\myparagraph{Adapting \pbe for non-experts}
Measuring the usability of \pbe with non-experts is a natural direction
for future research. Additionally, we make the following recommendations
for tasks that may be performed differently for non-experts.

\emparagraph{1. Collecting examples:} To ease the burden of creating
tags, non-experts may be provided with a large and diverse collection of
tags (e.g., the 40 tags obtained in our study) as a baseline for
specifying examples. Further, usable interfaces may be considered for
non experts (e.g., ``interactive dropdowns''~\cite{jkkg10a,jkkg10b} to
collect examples).

\emparagraph{2. Collecting Weights:} Collecting weights from non-experts
is another challenge for future work. Future work may consider using
visual ``sliders'' for weight collection, for precise and usable weight
assignments.

\emparagraph{3. System Integration:} \pbe may be integrated into
existing systems that protect user-specific data from disclosure to the
network (e.g., Weir~\cite{naej16} and Aquifer~\cite{ne13}). On such
systems, users may want to override policy predictions by \pbe at
runtime, or provide feedback, requiring a trusted path between the user
and \pbe.  A feedback mechanism may also improve future predictions.

\myparagraph{Adapting to Change}
Another direction for future work is detecting potential change in the
user's policy. While detecting change may be impossible without external
input in some cases, there is value in evaluating solutions in other
cases. Lessons from prior work that measures policy
changes for file access control may be used to determine the causes for
policy change~\cite{sg09}. Persuasive technologies designed by prior
research may also provide ways to encourage the user to report change
when it happens~\cite{cml09,mc12,pgb+10}. Future work
may also be directed at predicting which example or tag is likely to
change, using existing information (e.g., weights, frequency in
examples). Our intuition is that strategies used for cache replacement
(e.g., least recently used or LRU~\cite{sj94}) may apply, at least as a
starting point.

Finally, active learning may also be used to suggest new examples to the
user, which is not the focus of this paper, and lies in the broader
scope for future work.

\vspace{-1em}
\section{Threats to Validity}
\label{sec:discussion} 
\vspace{-1em}
In this paper, we provide a general framework for specifying policies
for user-specific data. Individual aspects of our framework may be
iteratively refined in the future.  
We identify specific limitations of the current state of our
approach and its evaluation as follows:

We evaluate feasibility with expert users.  While our participants
provide a significant number of policy examples, the number of
participants is small, hence we cannot generalize to the
broader user population. However, because this specialized set of users
are likely to have more complex policies than most users, we view our
feasibility study as a sufficient ``stress test'' of \pbe.

Additionally, since even expert users can make bad security
decisions~\cite{yho+14}, our expert-specified policy examples are not
expected to be error-free.  Indeed, we use our interactive approach to
help users find potential errors.

Our policy scenario is described as a conjunction of variables
(Section~\ref{sec:policy-example}). While it is easy to see how such a
format may generalize to any policy that may be expressed as a
conjunction of data objects or conditions, a thorough evaluation of
expressibility may be required. 

Finally, in Section~\ref{sec:active-learning}, we propose a simple
greedy approach to satisfy graph invariants. A more complex approach
(e.g., using dynamic programming) may be integrated into \pbe without
any significant changes.

\vspace{-1em}
\section{Conclusion} 
\label{sec:conc}
\vspace{-0.5em}

We introduced the paradigm of Policy by Example (\pbe) for user-specific
policy specification. \pbe enables users to express data-use scenarios
in policy examples, and predicts policy decisions for new scenarios.
In our feasibility study, \pbe demonstrated better prediction
performance than naive approaches. A key contribution of \pbe is its
active learning approach for engaging users in finding and potentially
incorrect policy decisions in their examples, which we demonstrated to
be five times as effective as manual reviews. Finally, we analyzed our
incorrect predictions and learned lessons that motivate future research
in this promising new domain.

{\footnotesize
\bibliographystyle{acm}
\bibliography{os,enck,add_to_os,phone,pl,policy,misc,ml,privacy,usability}
}
\appendix

\section{Tags used in the User Study}
\label{app:tags}
\begin{figure*}[t]
    \centering
    \includegraphics[width=6.5in]{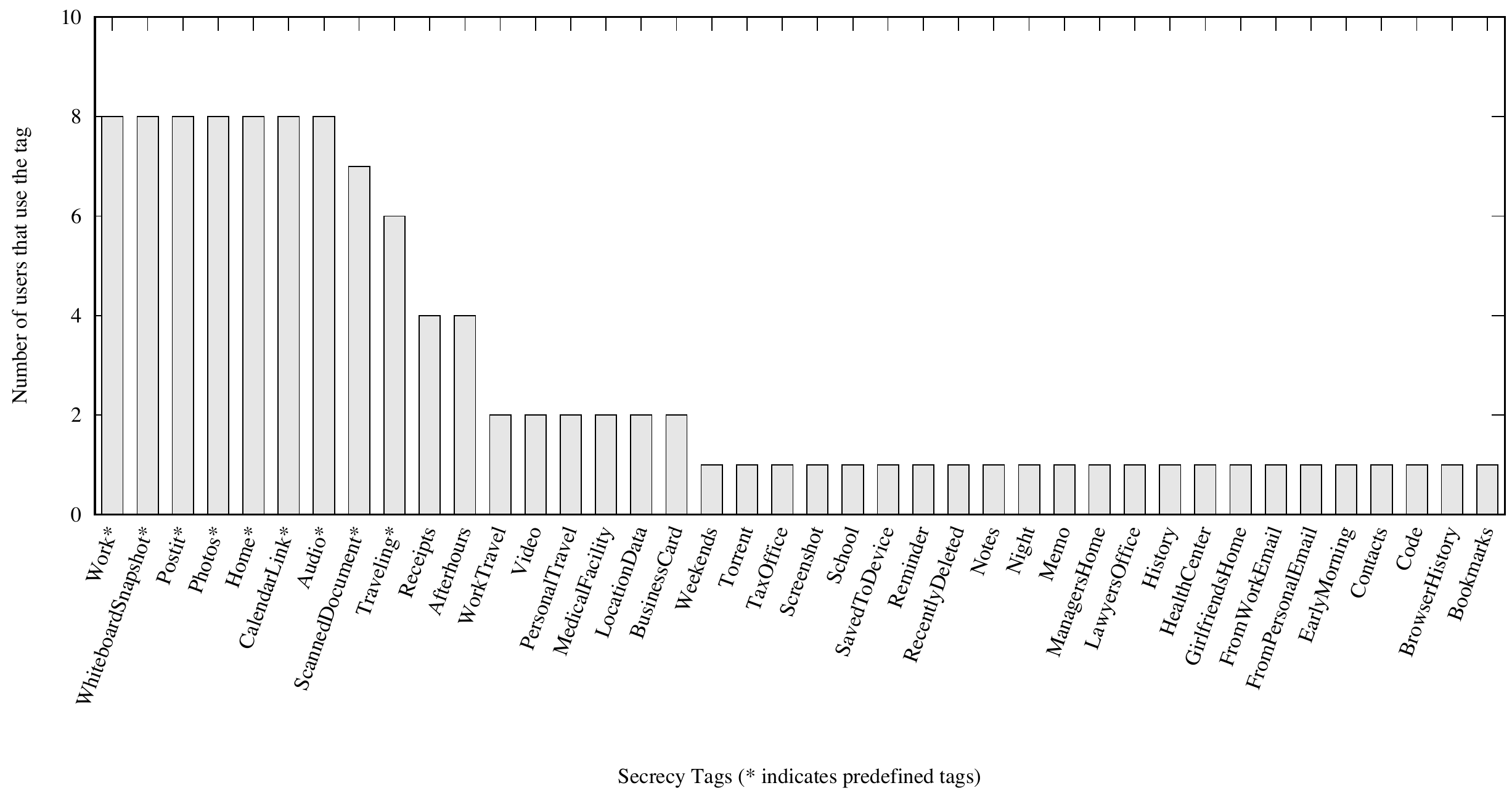}
    \caption{Tags used in the user study, and the number of users (out of 8) that use each tag.}
    \label{fig:tags}
\end{figure*}

Figure~\ref{fig:tags} shows the tags used in this study. We provided 9
tags, while the rest were created by users.

\section{Feasibility Study details}
\label{app:feasibility_details}

The data collection and experiments were performed using semi-structured
interviews. Most tasks (i.e., collecting examples, reviewing examples,
and testing) took about 75 minutes, whereas collecting weights took
about 20 minutes on average.  While breaks were offered as a part of the
experimental design, no participant elected to take their break.

\subsection{Collecting Examples}
\label{app:examples_script}
\begin{itemize}
  \item In this task, you will provide context-policy examples.
  \item You will be given a list of predefined context tags. You can use
    0 or more of these tags, and also create your own tags.
  \item You may combine tags to describe the context of a scenario. You
    will then be required to indicate a policy decision (i.e., 0 for
    deny and one for allow) for the policies provided.
  \item Each line on the example sheet has space for the context (i.e.,
    combination of tags), and a column for each policy.
\end{itemize}
\subsection{Collecting Weights}
\label{app:weights_script}
This phase consisted of two tasks, grouping tags and ordering groups. We
provide the instructions given to users as follows:
\subsubsection{Grouping tags} 
\begin{figure*}[t]
    \centering
    \includegraphics[width=6in]{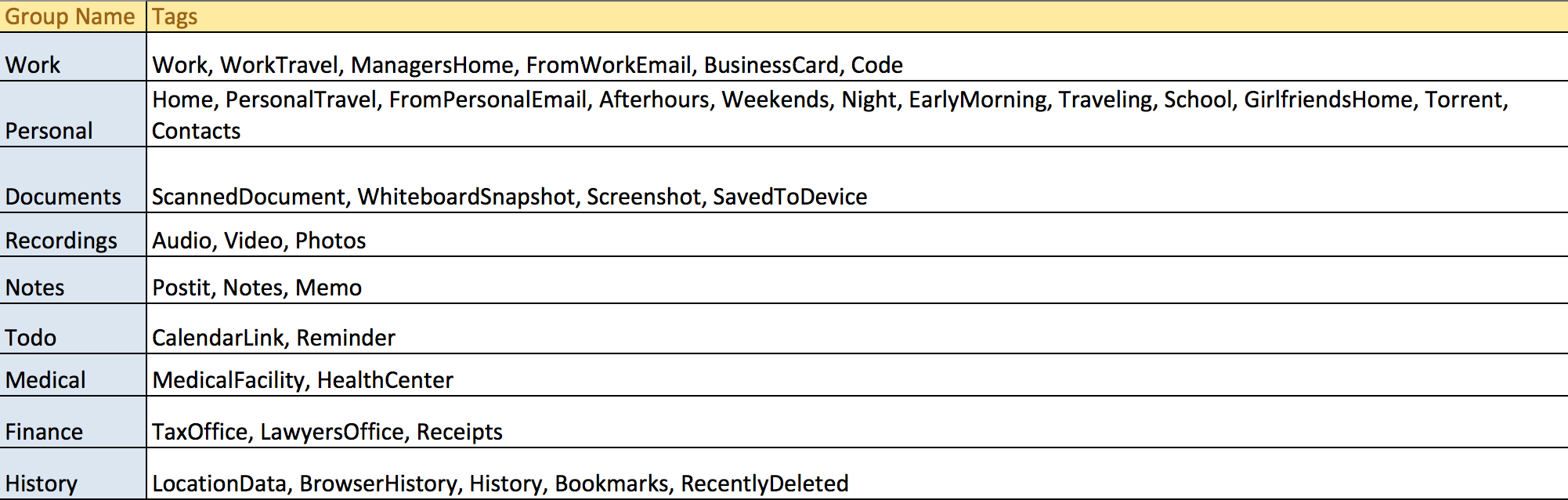}
    \caption{Screenshot of the tag groups customized by \user{1}}
    \label{fig:tag_groups}
\end{figure*}

We provide an example of customized tag-group memberships in
Figure~\ref{fig:tag_groups}. 
\begin{figure}[t]
    \centering
    \includegraphics[width=1.2in]{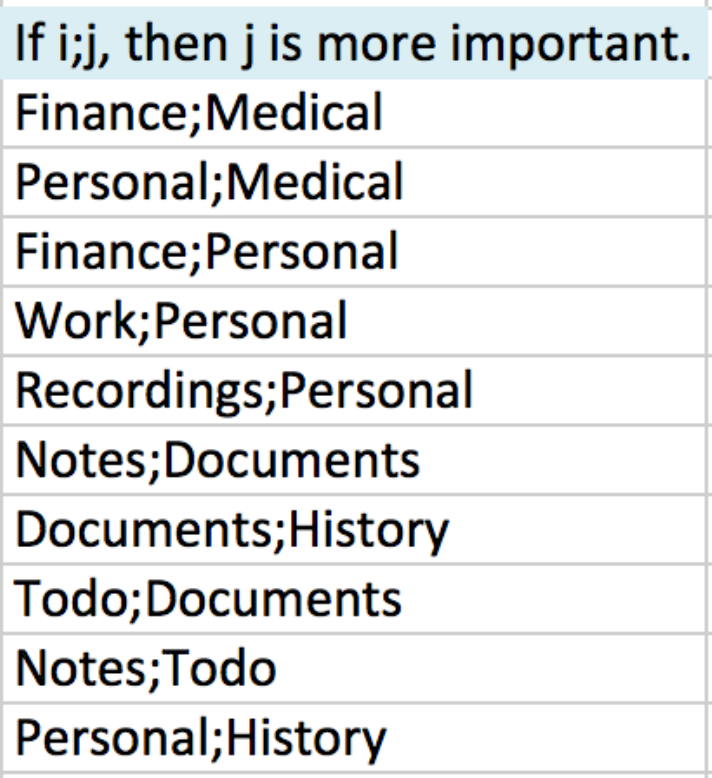}
    \vspace{-0.5em}
    \caption{Screenshot of \user{1}'s ordering of tag groups.}
    \label{fig:p1_order}
    \vspace{-1em}
\end{figure}
\begin{itemize}
  \item In this task, you will receive a spreadsheet containing groups,
    and information tags included in those groups. Note that the tags
    may include not only the tags you defined in the initial interview,
    but also tags defined by other users. 
  \item We have grouped tags that seem to be dealing with data of
    similar secrecy value. For instance, all the work-related tags such
    as ``Work'', ``WorkTravel'' are in the group ``Work''.
  \item The task is to verify group memberships, such that every tag
    belongs to the correct group as per your understanding.
  \item You can move tags around, i.e., remove them from one group, and
    add to another, but you cannot add, remove, or rename the groups
    themselves.
  \item The spreadsheet will also include a comments column, if you want
    to make a comment about a specific group, although comments are not
    required.
  \item Finally, the spreadsheet will include descriptions for some
    group names for reference.
\end{itemize}

\subsubsection{Ordering Groups}
\begin{itemize}
    \item In this task, you will be given a set of partial order relations
    among the groups described in the previous experiment. This set of
    relations is only a baseline. 
    \item A relation between groups A and B, such that A;B, means that B
    is more “important” than A. Since our policies are information
    secrecy-related,  “more important” may be understood as “more
    sensitive”.
  \item Your task is to modify (i.e., add or remove) the given list of
    relations, i.e., to customize the orders according to the your data
    secrecy/privacy preferences.
  \item You will also be allowed to use your initial group assignment
    for reference.
  \item It is possible that the ordering of groups may be different for
    different policies. Therefore, you will be able to use different
    sets of relations for different policies (total 7 policies). Please
    indicate if you want to use the same order for all policies, or if
    you would prefer to use different orders. This choice can be made or
    modified at any point of time throughout this task.
  \item In the end, I will confirm each order, for each policy (if the
    user chooses to have different orders for different policies). For
    instance, if the user enters A;B, I will confirm ``is B more
    important than A?''.
  \item Finally, you have the option of continuing to the next
    experiment after a break of 5-10 minutes, or calling it a day. 
\end{itemize}

\subsection{Review of Examples}
\label{app:review_script}
The review phase consisted of two tasks, namely a manual review, and a
semi-automated review using active learning. This section provides the scripts for
both tasks.

\subsubsection{Manual Review}
\begin{itemize}
  \item In this task, you will receive a set of spreadsheets (one per
policy) containing your examples (i.e., the context label + policy
decision) for that policy. 
  \item This is an opportunity for you to review your examples, and modify
the policy decision if necessary.  The context labels cannot be
modified.
  \item For each change you make, you will then indicate the cause of the
change in the respective column of one of the following hints:
  \begin{itemize}
      \item ``I have changed my mind'': i.e., my policy preferences have changed.
      \item ``It seems I made an error before''
      \item ``I don’t understand this policy example'': This could
	happen if you do not remember why you specified the policy, or
	are having trouble expressing it with tags you provided/used.
    \end{itemize}
  \item Finally, for each change you make, please provide justification in
the last column. This column may also be used for reasons other than the
said hints.
\item The investigator will go through the changes, and may ask you to
provide any missing justifications or causes.
\end{itemize}

\subsubsection{Semi-automatic Review}
\begin{itemize}
  \item In this task, our algorithm will suggest policy decisions for
  existing context labels that you have previously provided.
  \item  You must either agree (y) or disagree (n) to the decision. You can
  also skip by entering n twice.
  \item For every decision that you disagree to, please provide a short
  justification.
  \item For example, the algorithm may suggest “Denial of export to the
  WorkCloud when the data object with the context {created at Home,
  Photo}”. This suggestion will be presented as follows: 
  {\tt Home+Photos, WorkCloud = DENY (y/n)?}
  \item If you agree, the algorithm will make another suggestion, or stop.
  \item If you disagree, the algorithm will provide a text input for the
  justification.
  \item The task will consist of at most 15 questions.
\end{itemize}

\subsection{Testing with Random Examples}
\label{app:testing_script}

In this section, we describe the script for the phase of testing with
random examples. This phase was split into two tasks as well, i.e., the
task of labeling samples, and of the post test review.

\subsubsection{Labeling Test Samples}
\begin{itemize}
  \item In this task, you will receive a set of spreadsheets (one per
    policy) containing examples (i.e., the context). 
  \item Your task is to label the policy decision (allow/deny/I don’t
    know) for each example. 
\end{itemize}

\subsubsection{Post-test Review}
\begin{itemize}
    \item In this task, our algorithm will ask you to confirm policy
    decisions that you have previously provided.
    \item Please agree (y) or disagree (n) with your decision.
    \item For every decision that you agree to, please provide a short
      justification.
\end{itemize}

\section{Datasets}
\label{app:datasets}
The number of examples per participant in the specification and testing
datasets are shown in Tables~\ref{tbl:dataset}
and~\ref{tbl:testing_dataset} respectively.
\begin{table}[t]
\centering
\scriptsize
\caption{Number of unique example scenarios created by each user, as well
as the total number of policy examples created after assigning decisions
for 5 policies.}
\label{tbl:dataset}
\begin{tabular}{r|c|c}
  \Xhline{2\arrayrulewidth}
 {\bf Users}	& {\bf Example scenarios created}	& {\bf Policy examples}\\ 
        ~ 	& {\bf ($n$)} 	 	& {\bf for 5 policies ($n*5$)} \\
  \Xhline{2\arrayrulewidth}
  \user{1} & 52 & 260\\
  \user{2} & 23 & 115\\
  \user{3} & 25 & 125\\
  \user{4} & 24 & 120\\
  \user{5} & 33 & 165\\
  \user{6} & 26 & 130\\
  \user{7} & 21 & 105\\
  \user{8} & 42 & 210\\
  \Xhline{2\arrayrulewidth}
  {\bf Total} & 246 & 1,230\\
  \Xhline{2\arrayrulewidth}
  \end{tabular} 
\end{table}

\begin{table}[t]
\centering
\scriptsize
\caption{Number of random policy scenarios created for testing
predictions per participant. Participants provide policy decisions (and
\pbe predicts) for 5 policies.}
\label{tbl:testing_dataset}
\begin{tabular}{r|c|c}
  \Xhline{2\arrayrulewidth}
 {\bf Users}	& {\bf Random Test Scenarios}    & {\bf Labeled Test examples}\\ 
        ~ 	& {\bf ($n$)} 	 	& {\bf for 5 policies ($n*5$)} \\
  \Xhline{2\arrayrulewidth}
  \user{1} & 26 & 130\\
  \user{2} & 11 & 55\\
  \user{3} & 12 & 60\\
  \user{4} & 12 & 60\\
  \user{5} & 16 & 80\\
  \user{6} & 14 & 70\\
  \user{7} & 10 & 50\\
  \user{8} & 21 & 105\\
  \Xhline{2\arrayrulewidth}
  {\bf Total} & 122 & 610\\
  \Xhline{2\arrayrulewidth}
  \end{tabular} 
  \vspace{-1em}
\end{table}

\section{Detailed split of causes of errors}
\label{app:analysis}

A detailed split of the causes of error, across users and policies, can
be seen in Table~\ref{tbl:incorrect}.
\begin{table*}
\scriptsize
\centering
\caption{Breakdown of incorrect predictions into misconfigured
weights~(W), policy change~(C), unconfirmed policy change~(U) and label
confusion~(L), across all participants and policies.}
\label{tbl:incorrect}
\setlength{\tabcolsep}{3pt}
\begin{tabular}{c|cccc|cccc|cccc|cccc|cccc|cccc|cccc|cccc}
  \Xhline{2\arrayrulewidth}
  ~                 & \multicolumn{4}{c|}{\bf \user{1}} & \multicolumn{4}{c|}{\bf \user{2}} & \multicolumn{4}{c|}{\bf \user{3}} & \multicolumn{4}{c|}{\bf \user{4}} & \multicolumn{4}{c|}{\bf \user{5}} & \multicolumn{4}{c|}{\bf \user{6}} & \multicolumn{4}{c|}{\bf \user{7}} & \multicolumn{4}{c}{\bf \user{8}} \\
{\bf Policy}	& {\bf W} & {\bf C} & {\bf U} &{\bf L} & {\bf W} & {\bf C} & {\bf U} &{\bf L} & {\bf W} & {\bf C} & {\bf U} &{\bf L} & {\bf W} & {\bf C} & {\bf U} &{\bf L} & {\bf W} & {\bf C} & {\bf U} &{\bf L} & {\bf W} & {\bf C} & {\bf U} &{\bf L} & {\bf W} & {\bf C} & {\bf U} &{\bf L} & {\bf W} & {\bf C} & {\bf U} &{\bf L} \\
  \Xhline{2\arrayrulewidth}
    WorkCloud	    	& 1&0&0&0 & 2&1&0&0	& 0&1&3&0 & 2&0&0&0	& 6&0&0&1 & 0&1&0&0	& 0&1&0&0	& 0&3&3&1\\
    PersonalCloud  	& 6&0&0&0 & 2&0&2&1	& 0&0&0&0 & 3&0&0&0	& 3&0&0&1 & 7&0&0&0 & 0&6&0&0	& 2&0&1&3\\
    WorkEmailApp   	& 1&0&0&0 & 2&2&1&0	& 0&0&2&0 & 2&0&0&0	& 5&0&0&1 & 0&2&0&0	& 0&1&0&0	& 1&2&0&0\\
    PersonalEmailApp   	& 6&0&0&0 & 2&0&2&1	& 0&0&0&0 & 3&0&0&0	& 4&1&0&1 & 7&0&0&0	& 0&5&0&0	& 1&1&1&1\\
    SocialApp   	& 4&1&0&0 & 2&0&0&1	& 3&0&0&0 & 1&0&0&0	& 1&0&0&0 & 0&0&0&0	& 0&0&0&0	& 0&2&0&0\\
  \Xhline{2\arrayrulewidth}
    Total 	       & 18&1&0&0 & 10&3&5&3 & 3&1&5&0 & 11&0&0&0 & 19&1&0&4 & 14&3&0&0 & 0&13&0&0 & 4&8&5&5\\
  \Xhline{2\arrayrulewidth}
  \end{tabular} 
\vskip2pt
\begin{flushleft}
  {\footnotesize $^{*}$Five incorrect predictions for \user{8} (i.e., 3
  in {\em PersonalEmailApp} and 2 in {\em WorkEmailApp}) are not
  included in the table due to insufficient information.
}
\end{flushleft}
\vspace{-1.5em}
\end{table*}

\end{document}